\documentclass[aps,prl,twocolumn,superscriptaddress]{revtex4-1}
\usepackage{pdfpages}
\makeatletter
\AtBeginDocument{\let\LS@rot\@undefined}
\makeatother
\usepackage{amssymb}
\usepackage{amsmath}
\usepackage{graphicx}
\usepackage{xcolor, ulem}
\usepackage[version=3]{mhchem}
\usepackage{footnote}
\usepackage{bm}
\usepackage{dsfont}
\usepackage{xr-hyper}
\usepackage{hyperref}
\usepackage{cleveref}

\crefname{equation}{Eq.}{Eqs.}
\Crefname{equation}{Equation}{Equations}
\crefname{figure}{Fig.}{Figs.}
\Crefname{figure}{Figure}{Figures}
\crefname{section}{Sect.}{Sects.}
\Crefname{section}{Section}{Sections}
\crefname{table}{Table}{Tables}
\crefname{appsec}{Appendix}{Appendices}

\graphicspath{{Figures/}}
\definecolor{adpcolor}{rgb}{0.36,0.54,0.66}

\usepackage[cal=boondox]{mathalfa}
\DeclareMathOperator{\Tr}{Tr}

\begin{document}

\title{Variational Quantum Simulation of Ultrastrong Light-Matter Coupling}
\author{Agustin Di Paolo}
\affiliation{Institut quantique and D\'epartement de Physique, Universit\'e de Sherbrooke, Sherbrooke J1K 2R1 QC, Canada}
\author{Panagiotis Kl. Barkoutsos}
\affiliation{IBM Research GmbH, Zurich Research Laboratory, S\"aumerstrasse 4, 8803 R\"uschlikon, Switzerland.}
\author{Ivano Tavernelli}
\affiliation{IBM Research GmbH, Zurich Research Laboratory, S\"aumerstrasse 4, 8803 R\"uschlikon, Switzerland.}
\author{Alexandre Blais}
\affiliation{Institut quantique and D\'epartement de Physique, Universit\'e de Sherbrooke, Sherbrooke J1K 2R1 QC, Canada}
\affiliation{Canadian Institute for Advanced Research, Toronto, ON, Canada}

\date{\today}

\begin{abstract}
We propose the simulation of quantum-optical systems in the ultrastrong-coupling regime using a variational quantum algorithm. More precisely, we introduce a short-depth variational form to prepare the groundstate of the multimode Dicke model on a quantum processor and present proof-of-principle results obtained via cloud access to an IBM device. We moreover provide an algorithm for characterizing the groundstate by Wigner state tomography. Our work is a first step towards digital quantum simulation of quantum-optical systems with potential applications to the spin-boson, Kondo and Jahn-Teller models.
\end{abstract}

\maketitle

Quantum simulation is one of the most prominent applications of quantum processors for solving problems in quantum physics and chemistry. Importantly, quantum simulation aims to circumvent the limited capabilities of classical computers to represent quantum states in exponentially large Hilbert spaces. Recently, a hybrid, quantum-classical simulation paradigm exploiting quantum variational principles has been introduced \cite{peruzzo2014variational}. Following this pioneering work, many other realizations of what is known as Variational Quantum Algorithm (VQA) have appeared in the literature \cite{o2016scalable,kandala2017hardware,colless2018computation,hempel2018quantum}. 

VQAs have been shown to have some robustness against noise and thus appear appropriate for the current generation of Noise-Intermediate-Scale-Quantum (NISQ) processors \cite{mcclean2016theory, colless2018computation, Kandala2019Error}. Although considerable effort has been devoted to solving proof-of-principle instances of problems in quantum chemistry \cite{o2016scalable,kandala2017hardware,colless2018computation,hempel2018quantum} and optimization \cite{moll2018quantum}, the general applicability of this approach to other domains in physics is still a subject of debate and interest \cite{reiner2016emulating,dumitrescu2018cloud}. Here, we use a VQA to simulate strongly interacting light-matter models. In particular, we focus on obtaining the groundstate of a set of two-level atoms coupled to electromagnetic modes, which is of fundamental interest and has practical applications for example for quantum-information processing and sensing \cite{ciuti2006input,ashhab2010qubit,beaudoin2011dissipation,kockum2019ultrastrong,RevModPhys.91.025005}.  

The simplest case corresponds to that of a two-level atom coupled to a cavity mode and is described by the quantum Rabi Hamiltonian
\begin{equation}
H/\hbar = \frac{\omega_q}{2}\sigma^z + \omega_c a^{\dagger} a + g\sigma^x(a+a^{\dagger}).
\label{eq:quantum Rabi Hamiltonian}
\end{equation}
Here, $\omega_q$ and $\omega_c$ are the atomic and the electromagnetic-mode frequencies, $\sigma^{\mu}$ ($\mu=x,y,z$) the Pauli matrices and $a$ ($a^{\dagger}$) the annihilation (creation) operator for the oscillator, respectively. If the light-matter coupling constant, $g$, is small compared to the systems' frequencies, \cref{eq:quantum Rabi Hamiltonian} reduces to the Jaynes-Cummings Hamiltonian \cite{haroche2006exploring}. Under these conditions, the terms $\sigma^{+}a$ and $\sigma^{-}a^{\dagger}$, where $\sigma^{\pm}=(\sigma^{x}\pm i\sigma^y)/2$, lead to an exchange of a single excitation between the atom and the oscillator mode. Provided that $g$ is greater than the decoherence rates of the atom and the cavity, this regime of light-matter interaction is referred to as strong coupling, and it is widely exploited for quantum-information processing purposes \cite{blais2004cavity}.

As $g$ approaches a significant fraction of the bare atom and cavity frequencies, or becomes the largest energy scale in \cref{eq:quantum Rabi Hamiltonian}, the atom-cavity system enters the ultrastrong- (USC) and deep-strong-coupling (DSC) regimes, respectively \cite{ciuti2006input,bourassa2009ultrastrong,ballester2012quantum,kockum2019ultrastrong,RevModPhys.91.025005}. In these cases, the presence of the counter-rotating terms ($\sigma^{+}a^{\dagger}$ and $\sigma^{-}a$) in \cref{eq:quantum Rabi Hamiltonian} needs to be taken into account. Perturbation theory provides an accurate description for coupling strengths in the range of $10\%-30\%$ of the system's frequencies, but has limited applicability beyond that regime \cite{kockum2019ultrastrong}. While an exact analytical solution in principle exists for \cref{eq:quantum Rabi Hamiltonian} \cite{braak2011integrability}, larger systems involving multiple atoms and/or electromagnetic modes can only be handled numerically.

In the large-$g$ limit, however, the mean cavity-mode occupation number and its quantum fluctuations are large and a sizable Fock space is required for numerical simulations. The total Hilbert-space dimension can thus quickly become unpractical for many-particle systems. This fact motivates the search for powerful analytical and numerical methods \cite{ciuti2006input,hausinger2010qubit,ashhab2010qubit,beaudoin2011dissipation,diaz2016dynamical,shi2018ultrastrong,rivera2019variational} and quantum-simulation algorithms \cite{braumuller2017analog,ballester2012quantum,grimsmo2013cavity,kockum2019ultrastrong,RevModPhys.91.025005,andy2019} for this problem. 

We consider the generalization of \cref{eq:quantum Rabi Hamiltonian} to $N$ atoms and $M$ electromagnetic modes, given by 
\begin{equation}
H/\hbar = \sum_{i=1}^N \frac{\omega_{q i}}{2}\sigma^z_i + \sum_{k=1}^M\omega_k a^{\dagger}_k a_k + \sum_{i=1}^N\sum_{k=1}^M g_{ik}\sigma^x_i(a_k+a^{\dagger}_k),
\label{eq:Dicke Hamiltonian}
\end{equation}
where the constants $\{g_{ik}\}$ quantify the coupling strength between the $i^{\mathrm{th}}$ atom (of frequency $\omega_{q_i}$) and the $k^{\mathrm{th}}$ cavity mode (of frequency $\omega_k$)  referred below to as $k$-mode. For $M=1$, \cref{eq:Dicke Hamiltonian} reduces to the Dicke model, while the special case $N=1$ corresponds to the multimode quantum Rabi model. Digital quantum simulation of such models requires the encoding of the bosonic modes into qubit registers. We choose to use a Single-Excitation-Subspace (SES) encoding, in which the Fock space of a given $k$-mode is truncated to a maximum photon number $n_{k}^\mathrm{max}$, and represented by a qubit register of size $n_{k}^\mathrm{max}+1$ \cite{somma2003quantum,geller2015universal,barkoutsos2018quantum,avalle2014noisy}. A mapping from the $k$-mode Fock space to the single-excitation subspace of the qubit register is then defined as $|n_k\rangle\to|\tilde{n}_k\rangle = |0_0\dots 0_{n_k-1}1_{n_k}0_{n_k+1}\dots 0_{n_{k}^\mathrm{max}}\rangle$ for $n_k\in[0,n_{k}^\mathrm{max}]$, where the tilde is used hereafter to indicate encoded states and operators. Importantly, under SES encoding, quadratic bosonic Hamiltonians lead to next-neighbor interactions at most \cite{somma2003quantum}. Indeed, the $k$-mode annihilation operator maps to $a_k \to \tilde{a}_k= \sum_{{n}_k=0}^{{n}_{k}^\mathrm{max}-1}\sqrt{{n}_k+1}\sigma^{+}_{{n}_k}\sigma^{-}_{{n}_k+1}$, where $\sigma^{\pm}_{{n}_k}$ acts on the ${n}_k$th qubit of the $k$-mode register. The 2-local form of $\tilde{a}_k$ relaxes connectivity requirements on the $k$-mode qubit register and thus leads to a reduced gate count. Other encodings can be found in Refs.~\cite{macridin2018digital,macridin2018electron,andy2019}.

Finding the groundstate $|G\rangle$ of \cref{eq:Dicke Hamiltonian} by means of a VQA requires first to construct a proper variational form \cite{peruzzo2014variational,mcclean2016theory}. That is, a unitary $U(\bm{\theta})$ parametrized by a real-valued vector $\bm{\theta}$, such that
\begin{equation}
|\widetilde{G}\rangle\simeq U(\bm{\theta^*}) |\widetilde{\mathrm{vac}}\rangle,
\label{eq:variational_form_definition}
\end{equation}
where $|\widetilde{\mathrm{vac}}\rangle=|0_q\rangle\otimes_{k=1}^M|\widetilde{0}_k\rangle$ is the (encoded) noninteracting vacuum state, and $\bm{\theta^*}$ is obtained by classical minimization of the energy $E(\bm{\theta})=\langle\widetilde{\mathrm{vac}}|U^{\dagger}(\bm{\theta}) \widetilde{H} U(\bm{\theta})|\widetilde{\mathrm{vac}}\rangle$. Some intuition about a convenient choice of $U(\bm{\theta})$ can be gained from approximate disentangling transformations for \cref{eq:Dicke Hamiltonian} \cite{diaz2016dynamical,shi2018ultrastrong}. We refer to such transformations indistinctly as polaron Ans\"atze. The simplest transformation is obtained for the case of $N=1$, where it is useful to rotate $H\to H'=P^{\dagger}HP$ by means of a qubit-state-dependent displacement of the $k$-modes
\begin{equation}
P= \prod_{k=1}^M\exp[{g_k}\sigma^x(a_k + a_k^{\dagger})/(\omega_k + \omega'_q)],
\label{eq:polaron transformation multimode Rabi}
\end{equation}
where $\omega'_q$ is a renormalized frequency for the atom. As illustrated in Sect. IA of the Supplemental Material, the groundstate of $H'$ approaches the noninteracting groundstate of the atom-cavities system,  $|\mathrm{vac}\rangle$, in most coupling regimes. Therefore, the state $P|\mathrm{vac}\rangle$ approximates the groundstate $|G\rangle$ in the laboratory frame. 

Exploiting this fact to prepare $|\widetilde{G}\rangle$ on a quantum computer requires compiling $\widetilde{P}$ from single- and two-qubit gates, for instance, using a Trotter decomposition. The need for reducing the Trotter error, however, can lead to quantum circuits of large depth. Moreover, this approach is sensitive to errors arising from imperfect qubit control and noise. As a way around this problem, we propose to leverage the structure of the polaron transformation to obtain a short-depth variational from. We do this by parameterizing the Trotter decomposition of $\widetilde{P}$ and letting the variational algorithm adjust the unitary such that the groundstate-Ansatz energy is minimized. The variational form has not only the purpose of discovering short-depth quantum circuits for synthesizing the USC groundstate, but also to potentially improving on the disentangling capabilities of \cref{eq:polaron transformation multimode Rabi}. 

We construct the variational form by choosing a convenient Trotter decomposition of $\widetilde{P}$, first for the case of $N=1$. We introduce two $k$-mode operators, $\widetilde{X}_k^{e}$ and $\widetilde{X}_k^{o}$, which are defined such that $\widetilde{P}=\prod_{k=1}^M\exp[f_k\sigma^x(\widetilde{X}_k^{e} + \widetilde{X}_k^{o})]$, where $\{f_k=g_k/(\omega_k + \omega_q')\}$ is a set of constants that will latter play the role of variational parameters. Although $[\widetilde{X}_k^{e},\widetilde{X}_k^{o}]\neq 0$, $\widetilde{X}_k^{e}$ and $\widetilde{X}_k^{o}$ are respectively composed of commuting terms that act on even and odd sites of the $k$-mode qubit register (see the Supplemental Material, Sect. IB). The 2-local form of the encoded bosonic operators leads to an efficient implementation of the Trotter-expanded unitary
\begin{equation}
\widetilde{P}_{\bm{d}} \simeq \prod_{k=1}^M \prod_{s=1}^{d_k}\exp\Big(\frac{f_k}{d_k}\sigma^x\widetilde{X}_k^{e}\Big)\exp\Big(\frac{f_k}{d_k}\sigma^x\widetilde{X}_k^{o}\Big),
\label{eq:polaron transformation Trotter-expanded}
\end{equation}
where $d_k$ is the number of Trotter steps, that may vary with the $k$-mode index. As shown in Sect. IB of the Supplemental Material, the exponentials in this equation factorize exactly into a product of $n_k^{\mathrm{max}}$ controlled-exchange gates acting on next-neighbor qubits of the $k$-mode register with the atom register being the control qubit. The implementation of \cref{eq:polaron transformation Trotter-expanded} requires thus $n_k^{\mathrm{max}} \times d_k$ such gates per $k$-mode, adding to a total gate count of $\sum_{k=1}^M n_k^{\mathrm{max}} d_k$ before quantum-circuit compilation. This number grows linearly with the number of $k$-modes, their Fock-space dimension and the order of the Trotter expansion (Trotter depth). Interestingly, since \cref{eq:polaron transformation Trotter-expanded} parallelizes over the $k$-modes, its quantum-circuit depth does not scale with $M$.

For $N>1$, the resulting variational form incorporates blocks of the form of \cref{eq:polaron transformation Trotter-expanded} where the two-level-atom operator $\sigma^x\to\sigma_i^x$ is now labeled by $i\in[1,N]$ and alternated among the respective qubit registers (see the Supplemental Material, Sect. IC). This observation leads to the more general expression
\begin{equation}
\mathrm{Varform} = \prod_{i=1}^N \prod_{k=1}^M \prod_{s=1}^{d_{ik}}\exp\Big(\frac{f^s_{ik}}{d_{ik}}\sigma^x_i\widetilde{X}_k^{e}\Big)\exp\Big(\frac{f^s_{ik}}{d_{ik}}\sigma^x_i\widetilde{X}_{ik}^{o}\Big),
\label{eq:polaron transformation Trotter-expanded Dicke}
\end{equation}
where the coefficients $f_k\to f_{ik}^s$ are variational parameters that depend on the Trotter step $s\in[1,\dots,d_{ik}]$. Additionally, $f_{ik}^s$ can also be made a function of the $k$-mode photon number, such that $f_{ik}^s\to f_{ik}^s(n_k)$. As argued below, this trades shorter circuit depths for longer optimization runtime.

Important additional details apply, however, between the cases of $N=1$ and $N>1$. In particular, the case $N>1$ requires \cref{eq:polaron transformation Trotter-expanded Dicke} to be complemented by single-layer short-depth variational form that acts on the atoms' registers. This extra step initializes the polaron variational circuit to the state $|\widetilde{\mathrm{vac}}'\rangle = \prod_{k=1}^M|\psi_a\rangle|\tilde{0}_k\rangle$, where $|\psi_a\rangle$ is an entangled state of the atoms. The state $|\psi_a\rangle$ is determined by an auxiliary optimization loop specified in Sect. IC of the Supplemental Material. 

\cref{fig:1-atom} shows the results for the (a) single- and (b) two-mode Rabi Hamiltonian, and (c) single- and (d) two-mode Dicke model for $N=2$. The simulations assume the resonant case where atom and $k$-mode frequencies are set to $\omega_{q_i}=\omega_k\equiv\omega$ and $g_{ik}\equiv g$ is swept in $[0,\omega]$. The resonance condition leads to strong entanglement between the atoms and cavity modes due to the energetically favorable exchange of excitations. To quantify the performance of the variational form, we define the error metric $\Delta_{\mathrm{en}}=|(E_{\mathrm{vqe}}-E_{\mathrm{en}})/E_{\mathrm{en}}|$, accounting for the relative difference between the groundstate energy found by the VQA, $E_{\mathrm{vqe}}$, and the energy of the encoded groundstate, $E_{\mathrm{en}}$. An additional metric $\Delta_{\mathrm{ex}}=|(E_{\mathrm{en}}-E_{\mathrm{ex}})/E_{\mathrm{ex}}|$ quantifies the difference between $E_{\mathrm{en}}$ and the numerically exact groundstate energy. We evaluate $\Delta_{\mathrm{en}}$ and $\Delta_{\mathrm{ex}}$ as a function of $g/\omega$ for circuits with Trotter depth $d_{ik}=d$. The chosen Fock space truncation (see figure caption) leads to a small number of qubits while ensuring a relatively small $\Delta_{\mathrm{ex}}$. This choice seeks to reduce the quantum-hardware resources needed for simulation. 

\begin{figure} [t!]
\includegraphics[scale=0.9]{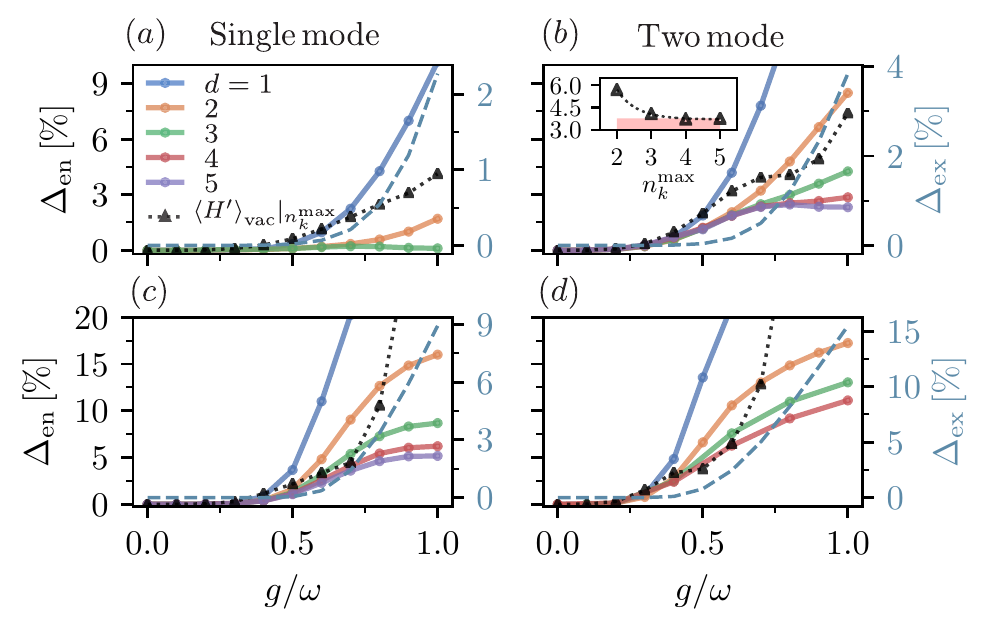}
\caption{\label{fig:1-atom} Groundstate-energy estimation for the single- and two-mode version of the Rabi and Dicke models in resonance conditions. Fock-space truncation in panels (a)-(d) is set to $n_k^{\mathrm{max}}=3$, $3$, $5$ and $4$ corresponding to $5$, $9$, $8$ and $12$ qubits, respectively. The legend is shared between all panels, although (a) and (d) display results only for $d_{i} < 4$ and $d_{ik} < 5$, respectively. We show the error metrics $\Delta_{\mathrm{en}}$ (left scale) and $\Delta_{\mathrm{ex}}$ (light-blue dashed line and right scale) defined in the main text, along with $\langle H'\rangle_{\mathrm{vac}}|_{\mathrm{n_k^{\mathrm{max}}}}$ (triangular markers). The inset in panel (b) shows $\langle H'\rangle_{\mathrm{vac}}|_{\mathrm{n_k^{\mathrm{max}}}}$ converging to a $4\%$ in the limit of large $n_k^{\mathrm{max}}$ (pink baseline) for $g/\omega=0.8$. This indicates that the minimum value of $\Delta_{\mathrm{en}}$ is reduced exponentially with $n_k^{\mathrm{max}}$, reaching an absolute lower bound determined by the entanglement capabilities of the polaron Ansatz. Simulations do not include noise and are done using Qiskit \cite{Qiskit}.}
\end{figure}

Results in panel (a) show that relative errors $\Delta_{\mathrm{en}}$ below $1\%$ are achieved by state-preparation circuits containing only 3 variational parameters ($d=3$). A similar accuracy is obtained for circuits with $d=2$ if additional parameters dependent on the $k$-mode photon-number are incorporated (not shown). Remarkably, for $d>1$, the energy of the variational Ansatz is significantly lower than $\langle H'\rangle_{\mathrm{vac}}|_{n_k^{\mathrm{max}}}$. The latter is the expectation value of \cref{eq:Dicke Hamiltonian} on the state $P|\mathrm{vac}\rangle|_{n_k^{\mathrm{max}}}$ within a truncated Fock space. This indicates that the variational algorithm can leverage the Trotter error to outperform the full polaron Ansatz under the same Fock-space restrictions and with very low circuit depth. Interestingly, we also find that the energy of the variational state falls below $\langle H'\rangle_{\mathrm{vac}}|_{n_k^{\mathrm{max}}\to\infty}$ in the full range of $g/\omega\in[0,1]$ (not shown). The error metric $\Delta_{\mathrm{ex}}$ remains below $\sim 2\%$ in all the cases. 

We observe a similar qualitative behavior for the two-mode simulations in panel (b), although $\Delta_{\mathrm{en}}$ increases to $\sim 2.5\%$ for $d=4$. The same accuracy is reached for circuits with $d=2$ when variational parameters for each $k$-mode photon number are introduced (not shown). We find that the accuracy limit is both due to finite Fock-space truncation errors and the disentangling capabilities of the polaron Ansatz. Increasing the number of two-level atoms in the model, while keeping the number of qubits of the order of $10$, leads to the results in panels (c)-(d) for which we find a maximum error of $\Delta_{\mathrm{en}}\simeq 5\%$ for $d=5$ in the first case, and of $\Delta_{\mathrm{en}}\simeq 8\%$ for $d=4$ in the second case. These results, however, are limited by Fock space truncation errors and can be improved by increasing the number of qubits in the simulations. It is worth noticing that, similarly to the case of $N=1$, these variational circuits outperform the polaron Ansatz significantly for the same conditions. 

The performance of the variational form may be improved further by means of simple modifications. For instance, a layer of a hardware-efficient (HE) gates \cite{kandala2017hardware} could be appended after each Trotter step, providing greater entangling capabilities for state preparation. Ideally, gates on such HE layers should conserve the number of excitations in the $k$-mode registers \cite{barkoutsos2018quantum}. Generalizations of \cref{eq:polaron transformation multimode Rabi} incorporating additional parameters are also a possibility \cite{chin2011generalized}. 

As the number of qubits scales with $\sim (n^{\mathrm{max}} + 1)^M$, simulating the performance of the proposed VQA on a classical computer becomes quickly expensive. Moreover, circuits of larger depth and number of qubits could likely benefit from quantum devices tailored to compile the polaron Ansatz in fewer gates. An option is to engineer the required controlled-two-qubit gates directly on the quantum hardware. Sect. V of the Supplemental Material illustrates such special-purpose devices in the context of circuit QED. 

\begin{figure}[t!]
\includegraphics[scale=1.]{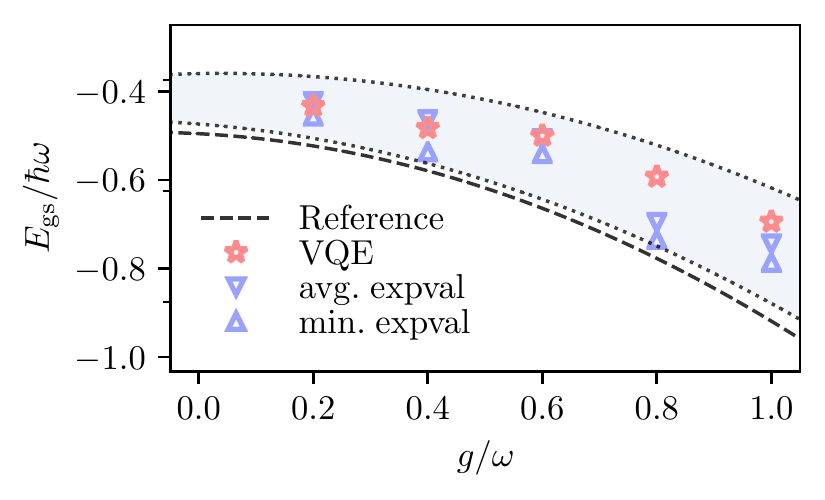}
\caption{\label{fig:real_hardware_plot} Variational quantum-optics simulation of the Rabi model in resonance conditions on a quantum processor. Shown is the groundstate energy as a function of the coupling strength, both in units of $\omega$. The cavity mode is encoded in a two-qubit register ($n_k^{\mathrm{max}}=1$). The light-blue bands enclose the range of results that are expected for 150 SPSA trials with levels of noise in the order of $0.1$ and $1.0$, relative to calibrated values (dotted lines). The black-dashed line is the encoded-groundstate energy. The star-shaped markers are the result of VQA runs for up to 150 SPSA trials on the quantum device. The pointing-up (pointing-down) triangular markers are the minimum (average) expectation value on quantum hardware of states that have been entirely optimized in the classical processor. The dispersion of such values is due to fluctuations in the level of noise of the quantum device between runs. Further details are provided in Sect. II of the Supplemental Material.}
\end{figure}

The results of \cref{fig:1-atom} suggest that the polaron variational form is a promising tool for investigating the USC groundstate in near-term quantum devices. For this reason, we implement the aforementioned strategy in currently available quantum hardware. Here, we use the IBM Q Poughkeepsie chip via the open-source framework Qiskit, taking advantage of the built-in SPSA optimizer \cite{spall1992multivariate, kandala2017hardware} and the readout error mitigation techniques of Qiskit-Ignis \cite{Qiskit}. We use three qubits for the quantum simulation, two of them encoding the bosonic mode. The groundstate energies found this way, shown in \cref{fig:real_hardware_plot} (star-shaped data points), are in good qualitative agreement with the theoretical estimations. 

We find that the main limitations on the accuracy of the VQA are due to the level of noise in the quantum processor and to the capabilities of the SPSA optimizer given a finite number of optimization steps. To investigate the effect of the latter against the former, we perform the VQA with a desktop computer, assuming a larger number of optimization steps and the calibrated noise model of the quantum hardware. This produces a set of variational states with optimal parametrization according to the classical simulation. We then evaluate the energy expectation value of such states on the quantum processor, performing mitigation of readout errors. The result of this experiment (triangular-shaped data points) reach better accuracies than those obtained by means of the hybrid quantum-classical VQA. This suggests that noise processes on the quantum hardware prevent high-accuracy solutions to be reached in a reasonable number of optimization steps via cloud access, in the order of $150$ SPSA trials. By controlling the level of noise in classical simulation, we also find that hybrid quantum-classical VQA solutions with $\Delta_{\mathrm{en}}\sim 1-2\%$ for $150$ SPSA trials are expected for noise levels one order of magnitude smaller than the present value. Note that in absence of noise, the number of optimizer steps required to reach numerical accuracy with respect to the reference value is very small in comparison, below $30$ in the entire $g/\omega\in[0,1]$ range. This allow us to conclude that the discrepancies encountered in the quantum-hardware runs are due the effect of noise and the limited optimizer calls rather than limitations of the proposed Ansatz.
% Higher accuracy estimations would thus result from improved quantum hardware, optimization and error-mitigation techniques \cite{mcclean2017hybrid,temme2017error,Kandala2019Error}. 

Following this proof-of-principle demonstration, we present an alternative method for characterizing the prepared groundstate. This technique could be useful to probe entanglement metrics and to distinguish between nearly degenerate states. The latter situation occurs, for instance, within the groundstate manifold of the quantum Rabi model approaching the DSC regime. To this end, we introduce the joint Wigner function for a set of $N$ qubits and $M$ bosonic modes as 
\begin{equation}
W_{\bm{l}}(\bm{\alpha})=\Tr[\rho\sigma_{1}^{l_1} \dots \sigma_{N}^{l_N}\,{2^M}\Pi(\bm{\alpha})/{\pi^M}],
\label{eq:join Wigner function}
\end{equation}
generalizing the definition given in Ref.~\cite{vlastakis2015characterizing} for the case of $N=M=1$. Here, $\{\sigma_{i}^{l_i},\,l_i\in[0,x,y,z]\}$ are the Pauli matrices for the $i^{\mathrm{th}}$ atom with $\sigma_i^{0}=\mathds{1}$. $\Pi(\bm{\alpha})=D({\bm{\alpha}})\Pi D^{\dagger}({\bm{\alpha}})$, where $\bm{\alpha} = (\alpha_1,\dots,\alpha_M)$, is a displaced joint-parity operator with $\Pi=\prod_{k=1}^M \exp(i\pi a_k^{\dagger}a_k)$ and $D({\bm{\alpha}})=\prod_{k=1}^M \exp(\alpha_k a^{\dagger}_k - \alpha^*_k a_k)$ for $\alpha_k\in\mathds{C}$. Inversion of \cref{eq:join Wigner function} gives the system's density matrix as $\rho={2^{M-N}}\sum_{\bm{l}}\int W_{\bm{l}}(\bm{\alpha}) \sigma_{1}^{l_1} \dots \sigma_{N}^{l_N} \Pi(\bm{\alpha})d^2\bm{\alpha}$, where the integral is performed over $d^2\bm{\alpha}=\prod_{k=1}^Md^2\alpha_k$ and the sum is extended to the $4^N$ possible values of $\bm{l}=(l_1,\dots,l_N)$. This relation can be used for state reconstruction \cite{vlastakis2015characterizing}.

Expanding \cref{eq:join Wigner function} in the Fock-state basis within the SES encoding we arrive to
\begin{equation}
\widetilde{W}_{\bm{l}}(\bm{\alpha})=\sum_{\bm{\tilde{n}}=0}^{\bm{\tilde{n}}^{\mathrm{max}}}(-1)^{\sum_{k=1}^M \tilde{n}_k}\Tr_q[2^M\Omega_{\bm{\tilde{n}}}(\bm{\alpha})\sigma_{1}^{l_1} \dots \sigma_{N}^{l_N}/\pi^M],
\label{eq:join Wigner function 2}
\end{equation}
where $\Tr_q$ is the trace operator over the atom registers and $\Omega_{\bm{\tilde{n}}} = \langle \tilde{n}_1\dots \tilde{n}_M| \widetilde{D}^{\dagger}(\bm{\alpha}) \tilde{\rho} \widetilde{D}(\bm{\alpha})|\tilde{n}_1\dots \tilde{n}_M\rangle$. $\widetilde{W}_{\bm{l}}(\bm{\alpha})$ can be sampled by executing a quantum circuit that performs the necessary state-tomography gates. While these gates are simply single-qubit rotations for the atom registers, tomography gates correspond to the application of $\widetilde{D}(\bm{\alpha})$ for the $k$-mode registers. Fortunately, the displacement operators can be easily implemented by a sequence of one- and two-qubit gates derived from a Trotter decomposition similar to that of the polaron transformation. Sect. III of the Supplemental Material includes further details.

\begin{figure} [t!]
\includegraphics[scale=1.]{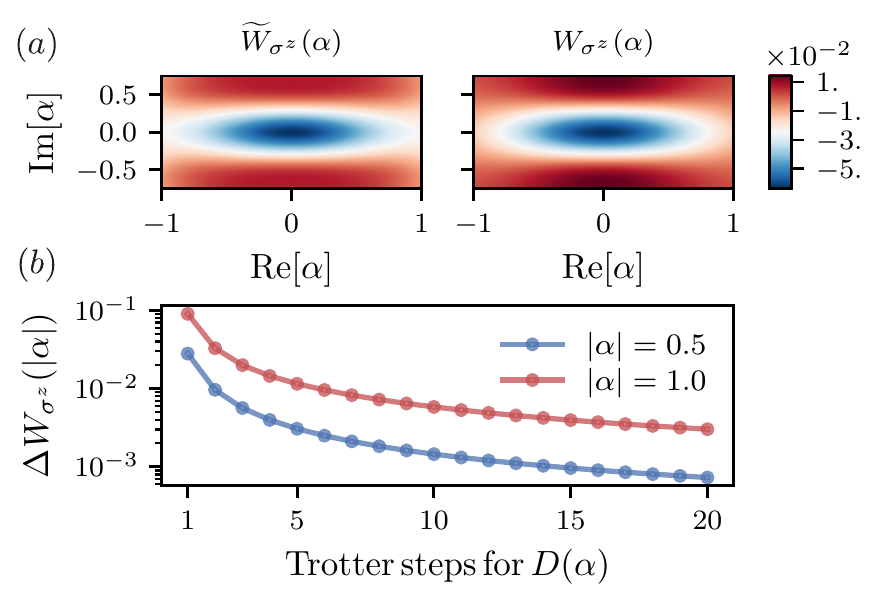}
\caption{\label{fig:wigner_Z} Reconstruction of the joint Wigner function of the Rabi model in resonance conditions and with $g/\omega=1$. (a) (Left) Sampled distribution, $\widetilde{W}_{\sigma^z}(\alpha)$, for an 8-qubit $k$-mode register and $2$ Trotter steps per imaginary and real components of $\widetilde{D}(\bm{\alpha})$. (Right) Numerically exact result $W_{\sigma^z}(\alpha)$. (b) Effect of the Trotter order of $D(\bm{\alpha})$ on the reconstructed distribution. The error metric is defined as $\Delta W_{\sigma^z}(|\alpha|) = \sqrt{\int_{|\alpha'|=0}^{|\alpha'|=|\alpha|} [\widetilde{W}_{\sigma^z}(\alpha')-{W}_{\sigma^z}(\alpha')]^2 d^2\alpha'}/\mathcal{N}_{|\alpha|}$, where $1/\mathcal{N}_{|\alpha|}$ is a normalization factor. The scaling of this metric with $|\alpha|$ follows the Trotter error, which scales as $|\alpha|^2/d$ with $d$ being the Trotter depth. For $|\alpha|$ fixed, the discrepancy between the two distributions saturates to a nonzero lower bound due to Fock-space truncation errors. The effect of noise has not been taken into account.}
\end{figure}

We demonstrate this approach numerically for the case of a single atom and a cavity mode in resonance with $g/\omega=1$. \cref{fig:wigner_Z} shows the reconstructed joint Wigner function $\widetilde{W}_{\sigma^z}({\alpha})$ for an 8-qubit $k$-mode register. The result is compared to the numerically exact distribution, which is not affected by Trotter or Fock space truncation errors. We observe a good qualitative agreement between the two distributions even for a few as $2$ Trotter steps. This agreement improves as the number of Trotter steps used to implement $D(\bm{\alpha})$ is increased, although the discrepancy between the two distributions remains bounded from below due to finite-dimensional encoding errors.

Finally, we discuss briefly the effect of common quantum error channels on the performance of the proposed VQA. It is worth highlighting that a SES encoding allows for damping errors of the form $|\tilde{n}_k\rangle\to|\widetilde{0}_k\rangle$ to be detectable by joint-parity measurements of the $k$-mode register. This enables postselection of uncorrupted states, which can significantly reduce the impact of noise on the proposed variational algorithm. However, the downsides of using a SES encoding reside in two main points. First, this encoding trades shorter quantum-circuit depths for a relatively large qubit overhead compared to other possible encodings \cite{macridin2018electron,macridin2018digital}. On the flip side, this compromise might be leveraged by current quantum processors, which are mostly limited by decoherence rather than by the number of qubits \cite{cross2018validating}. Second, noise channels that do not conserve the number of excitations in the $k$-mode qubit registers can become dominant for large qubit arrays (see Supplemental Material). This is a direct consequence of an exponential growth of the size of the complement of SES with the number of qubits in the simulation. We also note that the observations above are generic to other proposals using SES encodings \cite{somma2003quantum,avalle2014noisy,geller2015universal,barkoutsos2018quantum}.

In conclusion, we introduced a short-depth and few-parameter variational form to study the interacting light-matter groundstate of $N$ atoms and $M$ electromagnetic modes. We found that such a variational circuit can approximate the ultrastrong-coupling groundstate with very good accuracy. We implemented a proof-of-principle example on an IBM quantum processor, performing the mitigation of readout errors. Finally, we demonstrated the use of Wigner state-tomography to characterize the groundstate, and discussed the impact of noise on the variational algorithm. As the light-matter interaction Hamiltonian considered in this work is formally identical to the few-impurity spin-boson model, we envision applications to problems in condensed-matter physics for which the polaron transformation was originally introduced. The demonstration of quantum advantage by the variational approach introduced here is likely to require quantum-hardware of size and noise-resilience significantly beyond what is currently available. Our work is, however, a first step towards digital simulation of strongly interacting light-matter models with a quantum processor. 

\section*{Acknowledgments}
We thank Catherine Leroux, Andy C. Y. Li and David Poulin for insightful discussions and the members of the Qiskit-Aqua team, Richard Chen and Marco Pistoia, for support. PKB and IT acknowledge financial support from the Swiss National Science Foundation (SNF) through the grant No. 200021-179312. This work was undertaken thanks in part to funding from NSERC and the Canada First Research Excellence Fund.

IBM Q, Qiskit are trademarks of International Business Machines Corporation, registered in many jurisdictions worldwide. Other product or service names may be trademarks or service marks of IBM or other companies.

\bibliography{library}

\begin{thebibliography}{38}%
\makeatletter
\providecommand \@ifxundefined [1]{%
 \@ifx{#1\undefined}
}%
\providecommand \@ifnum [1]{%
 \ifnum #1\expandafter \@firstoftwo
 \else \expandafter \@secondoftwo
 \fi
}%
\providecommand \@ifx [1]{%
 \ifx #1\expandafter \@firstoftwo
 \else \expandafter \@secondoftwo
 \fi
}%
\providecommand \natexlab [1]{#1}%
\providecommand \enquote  [1]{``#1''}%
\providecommand \bibnamefont  [1]{#1}%
\providecommand \bibfnamefont [1]{#1}%
\providecommand \citenamefont [1]{#1}%
\providecommand \href@noop [0]{\@secondoftwo}%
\providecommand \href [0]{\begingroup \@sanitize@url \@href}%
\providecommand \@href[1]{\@@startlink{#1}\@@href}%
\providecommand \@@href[1]{\endgroup#1\@@endlink}%
\providecommand \@sanitize@url [0]{\catcode `\\12\catcode `\$12\catcode
  `\&12\catcode `\#12\catcode `\^12\catcode `\_12\catcode `\%12\relax}%
\providecommand \@@startlink[1]{}%
\providecommand \@@endlink[0]{}%
\providecommand \url  [0]{\begingroup\@sanitize@url \@url }%
\providecommand \@url [1]{\endgroup\@href {#1}{\urlprefix }}%
\providecommand \urlprefix  [0]{URL }%
\providecommand \Eprint [0]{\href }%
\providecommand \doibase [0]{http://dx.doi.org/}%
\providecommand \selectlanguage [0]{\@gobble}%
\providecommand \bibinfo  [0]{\@secondoftwo}%
\providecommand \bibfield  [0]{\@secondoftwo}%
\providecommand \translation [1]{[#1]}%
\providecommand \BibitemOpen [0]{}%
\providecommand \bibitemStop [0]{}%
\providecommand \bibitemNoStop [0]{.\EOS\space}%
\providecommand \EOS [0]{\spacefactor3000\relax}%
\providecommand \BibitemShut  [1]{\csname bibitem#1\endcsname}%
\let\auto@bib@innerbib\@empty
%</preamble>
\bibitem [{\citenamefont {Peruzzo}\ \emph {et~al.}(2014)\citenamefont
  {Peruzzo}, \citenamefont {McClean}, \citenamefont {Shadbolt}, \citenamefont
  {Yung}, \citenamefont {Zhou}, \citenamefont {Love}, \citenamefont
  {Aspuru-Guzik},\ and\ \citenamefont {O'brien}}]{peruzzo2014variational}%
  \BibitemOpen
  \bibfield  {author} {\bibinfo {author} {\bibfnamefont {A.}~\bibnamefont
  {Peruzzo}}, \bibinfo {author} {\bibfnamefont {J.}~\bibnamefont {McClean}},
  \bibinfo {author} {\bibfnamefont {P.}~\bibnamefont {Shadbolt}}, \bibinfo
  {author} {\bibfnamefont {M.-H.}\ \bibnamefont {Yung}}, \bibinfo {author}
  {\bibfnamefont {X.-Q.}\ \bibnamefont {Zhou}}, \bibinfo {author}
  {\bibfnamefont {P.~J.}\ \bibnamefont {Love}}, \bibinfo {author}
  {\bibfnamefont {A.}~\bibnamefont {Aspuru-Guzik}}, \ and\ \bibinfo {author}
  {\bibfnamefont {J.~L.}\ \bibnamefont {O'brien}},\ }\href@noop {} {\bibfield
  {journal} {\bibinfo  {journal} {Nature communications}\ }\textbf {\bibinfo
  {volume} {5}},\ \bibinfo {pages} {4213} (\bibinfo {year} {2014})}\BibitemShut
  {NoStop}%
\bibitem [{\citenamefont {O'Malley}\ \emph {et~al.}(2016)\citenamefont
  {O'Malley}, \citenamefont {Babbush}, \citenamefont {Kivlichan}, \citenamefont
  {Romero}, \citenamefont {McClean}, \citenamefont {Barends}, \citenamefont
  {Kelly}, \citenamefont {Roushan}, \citenamefont {Tranter}, \citenamefont
  {Ding} \emph {et~al.}}]{o2016scalable}%
  \BibitemOpen
  \bibfield  {author} {\bibinfo {author} {\bibfnamefont {P.~J.}\ \bibnamefont
  {O'Malley}}, \bibinfo {author} {\bibfnamefont {R.}~\bibnamefont {Babbush}},
  \bibinfo {author} {\bibfnamefont {I.~D.}\ \bibnamefont {Kivlichan}}, \bibinfo
  {author} {\bibfnamefont {J.}~\bibnamefont {Romero}}, \bibinfo {author}
  {\bibfnamefont {J.~R.}\ \bibnamefont {McClean}}, \bibinfo {author}
  {\bibfnamefont {R.}~\bibnamefont {Barends}}, \bibinfo {author} {\bibfnamefont
  {J.}~\bibnamefont {Kelly}}, \bibinfo {author} {\bibfnamefont
  {P.}~\bibnamefont {Roushan}}, \bibinfo {author} {\bibfnamefont
  {A.}~\bibnamefont {Tranter}}, \bibinfo {author} {\bibfnamefont
  {N.}~\bibnamefont {Ding}},  \emph {et~al.},\ }\href@noop {} {\bibfield
  {journal} {\bibinfo  {journal} {Physical Review X}\ }\textbf {\bibinfo
  {volume} {6}},\ \bibinfo {pages} {031007} (\bibinfo {year}
  {2016})}\BibitemShut {NoStop}%
\bibitem [{\citenamefont {Kandala}\ \emph {et~al.}(2017)\citenamefont
  {Kandala}, \citenamefont {Mezzacapo}, \citenamefont {Temme}, \citenamefont
  {Takita}, \citenamefont {Brink}, \citenamefont {Chow},\ and\ \citenamefont
  {Gambetta}}]{kandala2017hardware}%
  \BibitemOpen
  \bibfield  {author} {\bibinfo {author} {\bibfnamefont {A.}~\bibnamefont
  {Kandala}}, \bibinfo {author} {\bibfnamefont {A.}~\bibnamefont {Mezzacapo}},
  \bibinfo {author} {\bibfnamefont {K.}~\bibnamefont {Temme}}, \bibinfo
  {author} {\bibfnamefont {M.}~\bibnamefont {Takita}}, \bibinfo {author}
  {\bibfnamefont {M.}~\bibnamefont {Brink}}, \bibinfo {author} {\bibfnamefont
  {J.~M.}\ \bibnamefont {Chow}}, \ and\ \bibinfo {author} {\bibfnamefont
  {J.~M.}\ \bibnamefont {Gambetta}},\ }\href@noop {} {\bibfield  {journal}
  {\bibinfo  {journal} {Nature}\ }\textbf {\bibinfo {volume} {549}},\ \bibinfo
  {pages} {242} (\bibinfo {year} {2017})}\BibitemShut {NoStop}%
\bibitem [{\citenamefont {Colless}\ \emph {et~al.}(2018)\citenamefont
  {Colless}, \citenamefont {Ramasesh}, \citenamefont {Dahlen}, \citenamefont
  {Blok}, \citenamefont {Kimchi-Schwartz}, \citenamefont {McClean},
  \citenamefont {Carter}, \citenamefont {De~Jong},\ and\ \citenamefont
  {Siddiqi}}]{colless2018computation}%
  \BibitemOpen
  \bibfield  {author} {\bibinfo {author} {\bibfnamefont {J.~I.}\ \bibnamefont
  {Colless}}, \bibinfo {author} {\bibfnamefont {V.~V.}\ \bibnamefont
  {Ramasesh}}, \bibinfo {author} {\bibfnamefont {D.}~\bibnamefont {Dahlen}},
  \bibinfo {author} {\bibfnamefont {M.~S.}\ \bibnamefont {Blok}}, \bibinfo
  {author} {\bibfnamefont {M.}~\bibnamefont {Kimchi-Schwartz}}, \bibinfo
  {author} {\bibfnamefont {J.}~\bibnamefont {McClean}}, \bibinfo {author}
  {\bibfnamefont {J.}~\bibnamefont {Carter}}, \bibinfo {author} {\bibfnamefont
  {W.}~\bibnamefont {De~Jong}}, \ and\ \bibinfo {author} {\bibfnamefont
  {I.}~\bibnamefont {Siddiqi}},\ }\href@noop {} {\bibfield  {journal} {\bibinfo
   {journal} {Physical Review X}\ }\textbf {\bibinfo {volume} {8}},\ \bibinfo
  {pages} {011021} (\bibinfo {year} {2018})}\BibitemShut {NoStop}%
\bibitem [{\citenamefont {Hempel}\ \emph {et~al.}(2018)\citenamefont {Hempel},
  \citenamefont {Maier}, \citenamefont {Romero}, \citenamefont {McClean},
  \citenamefont {Monz}, \citenamefont {Shen}, \citenamefont {Jurcevic},
  \citenamefont {Lanyon}, \citenamefont {Love}, \citenamefont {Babbush} \emph
  {et~al.}}]{hempel2018quantum}%
  \BibitemOpen
  \bibfield  {author} {\bibinfo {author} {\bibfnamefont {C.}~\bibnamefont
  {Hempel}}, \bibinfo {author} {\bibfnamefont {C.}~\bibnamefont {Maier}},
  \bibinfo {author} {\bibfnamefont {J.}~\bibnamefont {Romero}}, \bibinfo
  {author} {\bibfnamefont {J.}~\bibnamefont {McClean}}, \bibinfo {author}
  {\bibfnamefont {T.}~\bibnamefont {Monz}}, \bibinfo {author} {\bibfnamefont
  {H.}~\bibnamefont {Shen}}, \bibinfo {author} {\bibfnamefont {P.}~\bibnamefont
  {Jurcevic}}, \bibinfo {author} {\bibfnamefont {B.~P.}\ \bibnamefont
  {Lanyon}}, \bibinfo {author} {\bibfnamefont {P.}~\bibnamefont {Love}},
  \bibinfo {author} {\bibfnamefont {R.}~\bibnamefont {Babbush}},  \emph
  {et~al.},\ }\href@noop {} {\bibfield  {journal} {\bibinfo  {journal}
  {Physical Review X}\ }\textbf {\bibinfo {volume} {8}},\ \bibinfo {pages}
  {031022} (\bibinfo {year} {2018})}\BibitemShut {NoStop}%
\bibitem [{\citenamefont {McClean}\ \emph {et~al.}(2016)\citenamefont
  {McClean}, \citenamefont {Romero}, \citenamefont {Babbush},\ and\
  \citenamefont {Aspuru-Guzik}}]{mcclean2016theory}%
  \BibitemOpen
  \bibfield  {author} {\bibinfo {author} {\bibfnamefont {J.~R.}\ \bibnamefont
  {McClean}}, \bibinfo {author} {\bibfnamefont {J.}~\bibnamefont {Romero}},
  \bibinfo {author} {\bibfnamefont {R.}~\bibnamefont {Babbush}}, \ and\
  \bibinfo {author} {\bibfnamefont {A.}~\bibnamefont {Aspuru-Guzik}},\
  }\href@noop {} {\bibfield  {journal} {\bibinfo  {journal} {New Journal of
  Physics}\ }\textbf {\bibinfo {volume} {18}},\ \bibinfo {pages} {023023}
  (\bibinfo {year} {2016})}\BibitemShut {NoStop}%
\bibitem [{\citenamefont {Kandala}\ \emph {et~al.}(2019)\citenamefont
  {Kandala}, \citenamefont {Temme}, \citenamefont {C{\'o}rcoles}, \citenamefont
  {Mezzacapo}, \citenamefont {Chow},\ and\ \citenamefont
  {Gambetta}}]{Kandala2019Error}%
  \BibitemOpen
  \bibfield  {author} {\bibinfo {author} {\bibfnamefont {A.}~\bibnamefont
  {Kandala}}, \bibinfo {author} {\bibfnamefont {K.}~\bibnamefont {Temme}},
  \bibinfo {author} {\bibfnamefont {A.~D.}\ \bibnamefont {C{\'o}rcoles}},
  \bibinfo {author} {\bibfnamefont {A.}~\bibnamefont {Mezzacapo}}, \bibinfo
  {author} {\bibfnamefont {J.~M.}\ \bibnamefont {Chow}}, \ and\ \bibinfo
  {author} {\bibfnamefont {J.~M.}\ \bibnamefont {Gambetta}},\ }\href@noop {}
  {\bibfield  {journal} {\bibinfo  {journal} {Nature}\ }\textbf {\bibinfo
  {volume} {567}},\ \bibinfo {pages} {491} (\bibinfo {year}
  {2019})}\BibitemShut {NoStop}%
\bibitem [{\citenamefont {Moll}\ \emph {et~al.}(2018)\citenamefont {Moll},
  \citenamefont {Barkoutsos}, \citenamefont {Bishop}, \citenamefont {Chow},
  \citenamefont {Cross}, \citenamefont {Egger}, \citenamefont {Filipp},
  \citenamefont {Fuhrer}, \citenamefont {Gambetta}, \citenamefont {Ganzhorn}
  \emph {et~al.}}]{moll2018quantum}%
  \BibitemOpen
  \bibfield  {author} {\bibinfo {author} {\bibfnamefont {N.}~\bibnamefont
  {Moll}}, \bibinfo {author} {\bibfnamefont {P.}~\bibnamefont {Barkoutsos}},
  \bibinfo {author} {\bibfnamefont {L.~S.}\ \bibnamefont {Bishop}}, \bibinfo
  {author} {\bibfnamefont {J.~M.}\ \bibnamefont {Chow}}, \bibinfo {author}
  {\bibfnamefont {A.}~\bibnamefont {Cross}}, \bibinfo {author} {\bibfnamefont
  {D.~J.}\ \bibnamefont {Egger}}, \bibinfo {author} {\bibfnamefont
  {S.}~\bibnamefont {Filipp}}, \bibinfo {author} {\bibfnamefont
  {A.}~\bibnamefont {Fuhrer}}, \bibinfo {author} {\bibfnamefont {J.~M.}\
  \bibnamefont {Gambetta}}, \bibinfo {author} {\bibfnamefont {M.}~\bibnamefont
  {Ganzhorn}},  \emph {et~al.},\ }\href@noop {} {\bibfield  {journal} {\bibinfo
   {journal} {Quantum Science and Technology}\ }\textbf {\bibinfo {volume}
  {3}},\ \bibinfo {pages} {030503} (\bibinfo {year} {2018})}\BibitemShut
  {NoStop}%
\bibitem [{\citenamefont {Reiner}\ \emph {et~al.}(2016)\citenamefont {Reiner},
  \citenamefont {Marthaler}, \citenamefont {Braum\"uller}, \citenamefont
  {Weides},\ and\ \citenamefont {Sch\"on}}]{reiner2016emulating}%
  \BibitemOpen
  \bibfield  {author} {\bibinfo {author} {\bibfnamefont {J.-M.}\ \bibnamefont
  {Reiner}}, \bibinfo {author} {\bibfnamefont {M.}~\bibnamefont {Marthaler}},
  \bibinfo {author} {\bibfnamefont {J.}~\bibnamefont {Braum\"uller}}, \bibinfo
  {author} {\bibfnamefont {M.}~\bibnamefont {Weides}}, \ and\ \bibinfo {author}
  {\bibfnamefont {G.}~\bibnamefont {Sch\"on}},\ }\href {\doibase
  10.1103/PhysRevA.94.032338} {\bibfield  {journal} {\bibinfo  {journal} {Phys.
  Rev. A}\ }\textbf {\bibinfo {volume} {94}},\ \bibinfo {pages} {032338}
  (\bibinfo {year} {2016})}\BibitemShut {NoStop}%
\bibitem [{\citenamefont {Dumitrescu}\ \emph {et~al.}(2018)\citenamefont
  {Dumitrescu}, \citenamefont {McCaskey}, \citenamefont {Hagen}, \citenamefont
  {Jansen}, \citenamefont {Morris}, \citenamefont {Papenbrock}, \citenamefont
  {Pooser}, \citenamefont {Dean},\ and\ \citenamefont
  {Lougovski}}]{dumitrescu2018cloud}%
  \BibitemOpen
  \bibfield  {author} {\bibinfo {author} {\bibfnamefont {E.~F.}\ \bibnamefont
  {Dumitrescu}}, \bibinfo {author} {\bibfnamefont {A.~J.}\ \bibnamefont
  {McCaskey}}, \bibinfo {author} {\bibfnamefont {G.}~\bibnamefont {Hagen}},
  \bibinfo {author} {\bibfnamefont {G.~R.}\ \bibnamefont {Jansen}}, \bibinfo
  {author} {\bibfnamefont {T.~D.}\ \bibnamefont {Morris}}, \bibinfo {author}
  {\bibfnamefont {T.}~\bibnamefont {Papenbrock}}, \bibinfo {author}
  {\bibfnamefont {R.~C.}\ \bibnamefont {Pooser}}, \bibinfo {author}
  {\bibfnamefont {D.~J.}\ \bibnamefont {Dean}}, \ and\ \bibinfo {author}
  {\bibfnamefont {P.}~\bibnamefont {Lougovski}},\ }\href@noop {} {\bibfield
  {journal} {\bibinfo  {journal} {Physical Review Letters}\ }\textbf {\bibinfo
  {volume} {120}},\ \bibinfo {pages} {210501} (\bibinfo {year}
  {2018})}\BibitemShut {NoStop}%
\bibitem [{\citenamefont {Ciuti}\ and\ \citenamefont
  {Carusotto}(2006)}]{ciuti2006input}%
  \BibitemOpen
  \bibfield  {author} {\bibinfo {author} {\bibfnamefont {C.}~\bibnamefont
  {Ciuti}}\ and\ \bibinfo {author} {\bibfnamefont {I.}~\bibnamefont
  {Carusotto}},\ }\href@noop {} {\bibfield  {journal} {\bibinfo  {journal}
  {Physical Review A}\ }\textbf {\bibinfo {volume} {74}},\ \bibinfo {pages}
  {033811} (\bibinfo {year} {2006})}\BibitemShut {NoStop}%
\bibitem [{\citenamefont {Ashhab}\ and\ \citenamefont
  {Nori}(2010)}]{ashhab2010qubit}%
  \BibitemOpen
  \bibfield  {author} {\bibinfo {author} {\bibfnamefont {S.}~\bibnamefont
  {Ashhab}}\ and\ \bibinfo {author} {\bibfnamefont {F.}~\bibnamefont {Nori}},\
  }\href@noop {} {\bibfield  {journal} {\bibinfo  {journal} {Physical Review
  A}\ }\textbf {\bibinfo {volume} {81}},\ \bibinfo {pages} {042311} (\bibinfo
  {year} {2010})}\BibitemShut {NoStop}%
\bibitem [{\citenamefont {Beaudoin}\ \emph {et~al.}(2011)\citenamefont
  {Beaudoin}, \citenamefont {Gambetta},\ and\ \citenamefont
  {Blais}}]{beaudoin2011dissipation}%
  \BibitemOpen
  \bibfield  {author} {\bibinfo {author} {\bibfnamefont {F.}~\bibnamefont
  {Beaudoin}}, \bibinfo {author} {\bibfnamefont {J.~M.}\ \bibnamefont
  {Gambetta}}, \ and\ \bibinfo {author} {\bibfnamefont {A.}~\bibnamefont
  {Blais}},\ }\href@noop {} {\bibfield  {journal} {\bibinfo  {journal}
  {Physical Review A}\ }\textbf {\bibinfo {volume} {84}},\ \bibinfo {pages}
  {043832} (\bibinfo {year} {2011})}\BibitemShut {NoStop}%
\bibitem [{\citenamefont {Kockum}\ \emph {et~al.}(2019)\citenamefont {Kockum},
  \citenamefont {Miranowicz}, \citenamefont {De~Liberato}, \citenamefont
  {Savasta},\ and\ \citenamefont {Nori}}]{kockum2019ultrastrong}%
  \BibitemOpen
  \bibfield  {author} {\bibinfo {author} {\bibfnamefont {A.~F.}\ \bibnamefont
  {Kockum}}, \bibinfo {author} {\bibfnamefont {A.}~\bibnamefont {Miranowicz}},
  \bibinfo {author} {\bibfnamefont {S.}~\bibnamefont {De~Liberato}}, \bibinfo
  {author} {\bibfnamefont {S.}~\bibnamefont {Savasta}}, \ and\ \bibinfo
  {author} {\bibfnamefont {F.}~\bibnamefont {Nori}},\ }\href@noop {} {\bibfield
   {journal} {\bibinfo  {journal} {Nature Reviews Physics}\ }\textbf {\bibinfo
  {volume} {1}},\ \bibinfo {pages} {19} (\bibinfo {year} {2019})}\BibitemShut
  {NoStop}%
\bibitem [{\citenamefont {Forn-D\'{\i}az}\ \emph {et~al.}(2019)\citenamefont
  {Forn-D\'{\i}az}, \citenamefont {Lamata}, \citenamefont {Rico}, \citenamefont
  {Kono},\ and\ \citenamefont {Solano}}]{RevModPhys.91.025005}%
  \BibitemOpen
  \bibfield  {author} {\bibinfo {author} {\bibfnamefont {P.}~\bibnamefont
  {Forn-D\'{\i}az}}, \bibinfo {author} {\bibfnamefont {L.}~\bibnamefont
  {Lamata}}, \bibinfo {author} {\bibfnamefont {E.}~\bibnamefont {Rico}},
  \bibinfo {author} {\bibfnamefont {J.}~\bibnamefont {Kono}}, \ and\ \bibinfo
  {author} {\bibfnamefont {E.}~\bibnamefont {Solano}},\ }\href {\doibase
  10.1103/RevModPhys.91.025005} {\bibfield  {journal} {\bibinfo  {journal}
  {Rev. Mod. Phys.}\ }\textbf {\bibinfo {volume} {91}},\ \bibinfo {pages}
  {025005} (\bibinfo {year} {2019})}\BibitemShut {NoStop}%
\bibitem [{\citenamefont {Haroche}\ and\ \citenamefont
  {Raimond}(2006)}]{haroche2006exploring}%
  \BibitemOpen
  \bibfield  {author} {\bibinfo {author} {\bibfnamefont {S.}~\bibnamefont
  {Haroche}}\ and\ \bibinfo {author} {\bibfnamefont {J.-M.}\ \bibnamefont
  {Raimond}},\ }\href@noop {} {\emph {\bibinfo {title} {Exploring the quantum:
  atoms, cavities, and photons}}}\ (\bibinfo  {publisher} {Oxford university
  press},\ \bibinfo {year} {2006})\BibitemShut {NoStop}%
\bibitem [{\citenamefont {Blais}\ \emph {et~al.}(2004)\citenamefont {Blais},
  \citenamefont {Huang}, \citenamefont {Wallraff}, \citenamefont {Girvin},\
  and\ \citenamefont {Schoelkopf}}]{blais2004cavity}%
  \BibitemOpen
  \bibfield  {author} {\bibinfo {author} {\bibfnamefont {A.}~\bibnamefont
  {Blais}}, \bibinfo {author} {\bibfnamefont {R.-S.}\ \bibnamefont {Huang}},
  \bibinfo {author} {\bibfnamefont {A.}~\bibnamefont {Wallraff}}, \bibinfo
  {author} {\bibfnamefont {S.~M.}\ \bibnamefont {Girvin}}, \ and\ \bibinfo
  {author} {\bibfnamefont {R.~J.}\ \bibnamefont {Schoelkopf}},\ }\href@noop {}
  {\bibfield  {journal} {\bibinfo  {journal} {Physical Review A}\ }\textbf
  {\bibinfo {volume} {69}},\ \bibinfo {pages} {062320} (\bibinfo {year}
  {2004})}\BibitemShut {NoStop}%
\bibitem [{\citenamefont {Bourassa}\ \emph {et~al.}(2009)\citenamefont
  {Bourassa}, \citenamefont {Gambetta}, \citenamefont {Abdumalikov~Jr},
  \citenamefont {Astafiev}, \citenamefont {Nakamura},\ and\ \citenamefont
  {Blais}}]{bourassa2009ultrastrong}%
  \BibitemOpen
  \bibfield  {author} {\bibinfo {author} {\bibfnamefont {J.}~\bibnamefont
  {Bourassa}}, \bibinfo {author} {\bibfnamefont {J.~M.}\ \bibnamefont
  {Gambetta}}, \bibinfo {author} {\bibfnamefont {A.~A.}\ \bibnamefont
  {Abdumalikov~Jr}}, \bibinfo {author} {\bibfnamefont {O.}~\bibnamefont
  {Astafiev}}, \bibinfo {author} {\bibfnamefont {Y.}~\bibnamefont {Nakamura}},
  \ and\ \bibinfo {author} {\bibfnamefont {A.}~\bibnamefont {Blais}},\
  }\href@noop {} {\bibfield  {journal} {\bibinfo  {journal} {Physical Review
  A}\ }\textbf {\bibinfo {volume} {80}},\ \bibinfo {pages} {032109} (\bibinfo
  {year} {2009})}\BibitemShut {NoStop}%
\bibitem [{\citenamefont {Ballester}\ \emph {et~al.}(2012)\citenamefont
  {Ballester}, \citenamefont {Romero}, \citenamefont {Garc{\'\i}a-Ripoll},
  \citenamefont {Deppe},\ and\ \citenamefont {Solano}}]{ballester2012quantum}%
  \BibitemOpen
  \bibfield  {author} {\bibinfo {author} {\bibfnamefont {D.}~\bibnamefont
  {Ballester}}, \bibinfo {author} {\bibfnamefont {G.}~\bibnamefont {Romero}},
  \bibinfo {author} {\bibfnamefont {J.~J.}\ \bibnamefont {Garc{\'\i}a-Ripoll}},
  \bibinfo {author} {\bibfnamefont {F.}~\bibnamefont {Deppe}}, \ and\ \bibinfo
  {author} {\bibfnamefont {E.}~\bibnamefont {Solano}},\ }\href@noop {}
  {\bibfield  {journal} {\bibinfo  {journal} {Physical Review X}\ }\textbf
  {\bibinfo {volume} {2}},\ \bibinfo {pages} {021007} (\bibinfo {year}
  {2012})}\BibitemShut {NoStop}%
\bibitem [{\citenamefont {Braak}(2011)}]{braak2011integrability}%
  \BibitemOpen
  \bibfield  {author} {\bibinfo {author} {\bibfnamefont {D.}~\bibnamefont
  {Braak}},\ }\href@noop {} {\bibfield  {journal} {\bibinfo  {journal}
  {Physical Review Letters}\ }\textbf {\bibinfo {volume} {107}},\ \bibinfo
  {pages} {100401} (\bibinfo {year} {2011})}\BibitemShut {NoStop}%
\bibitem [{\citenamefont {Hausinger}\ and\ \citenamefont
  {Grifoni}(2010)}]{hausinger2010qubit}%
  \BibitemOpen
  \bibfield  {author} {\bibinfo {author} {\bibfnamefont {J.}~\bibnamefont
  {Hausinger}}\ and\ \bibinfo {author} {\bibfnamefont {M.}~\bibnamefont
  {Grifoni}},\ }\href@noop {} {\bibfield  {journal} {\bibinfo  {journal}
  {Physical Review A}\ }\textbf {\bibinfo {volume} {82}},\ \bibinfo {pages}
  {062320} (\bibinfo {year} {2010})}\BibitemShut {NoStop}%
\bibitem [{\citenamefont {D{\`\i}az-Camacho}\ \emph {et~al.}(2016)\citenamefont
  {D{\`\i}az-Camacho}, \citenamefont {Bermudez},\ and\ \citenamefont
  {Garc{\'\i}a-Ripoll}}]{diaz2016dynamical}%
  \BibitemOpen
  \bibfield  {author} {\bibinfo {author} {\bibfnamefont {G.}~\bibnamefont
  {D{\`\i}az-Camacho}}, \bibinfo {author} {\bibfnamefont {A.}~\bibnamefont
  {Bermudez}}, \ and\ \bibinfo {author} {\bibfnamefont {J.~J.}\ \bibnamefont
  {Garc{\'\i}a-Ripoll}},\ }\href@noop {} {\bibfield  {journal} {\bibinfo
  {journal} {Physical Review A}\ }\textbf {\bibinfo {volume} {93}},\ \bibinfo
  {pages} {043843} (\bibinfo {year} {2016})}\BibitemShut {NoStop}%
\bibitem [{\citenamefont {Shi}\ \emph {et~al.}(2018)\citenamefont {Shi},
  \citenamefont {Chang},\ and\ \citenamefont
  {Garc{\'\i}a-Ripoll}}]{shi2018ultrastrong}%
  \BibitemOpen
  \bibfield  {author} {\bibinfo {author} {\bibfnamefont {T.}~\bibnamefont
  {Shi}}, \bibinfo {author} {\bibfnamefont {Y.}~\bibnamefont {Chang}}, \ and\
  \bibinfo {author} {\bibfnamefont {J.~J.}\ \bibnamefont
  {Garc{\'\i}a-Ripoll}},\ }\href@noop {} {\bibfield  {journal} {\bibinfo
  {journal} {Physical Review Letters}\ }\textbf {\bibinfo {volume} {120}},\
  \bibinfo {pages} {153602} (\bibinfo {year} {2018})}\BibitemShut {NoStop}%
\bibitem [{\citenamefont {Rivera}\ \emph {et~al.}(2019)\citenamefont {Rivera},
  \citenamefont {Flick},\ and\ \citenamefont {Narang}}]{rivera2019variational}%
  \BibitemOpen
  \bibfield  {author} {\bibinfo {author} {\bibfnamefont {N.}~\bibnamefont
  {Rivera}}, \bibinfo {author} {\bibfnamefont {J.}~\bibnamefont {Flick}}, \
  and\ \bibinfo {author} {\bibfnamefont {P.}~\bibnamefont {Narang}},\
  }\href@noop {} {\bibfield  {journal} {\bibinfo  {journal} {Physical Review
  Letters}\ }\textbf {\bibinfo {volume} {122}},\ \bibinfo {pages} {193603}
  (\bibinfo {year} {2019})}\BibitemShut {NoStop}%
\bibitem [{\citenamefont {Braum{\"u}ller}\ \emph {et~al.}(2017)\citenamefont
  {Braum{\"u}ller}, \citenamefont {Marthaler}, \citenamefont {Schneider},
  \citenamefont {Stehli}, \citenamefont {Rotzinger}, \citenamefont {Weides},\
  and\ \citenamefont {Ustinov}}]{braumuller2017analog}%
  \BibitemOpen
  \bibfield  {author} {\bibinfo {author} {\bibfnamefont {J.}~\bibnamefont
  {Braum{\"u}ller}}, \bibinfo {author} {\bibfnamefont {M.}~\bibnamefont
  {Marthaler}}, \bibinfo {author} {\bibfnamefont {A.}~\bibnamefont
  {Schneider}}, \bibinfo {author} {\bibfnamefont {A.}~\bibnamefont {Stehli}},
  \bibinfo {author} {\bibfnamefont {H.}~\bibnamefont {Rotzinger}}, \bibinfo
  {author} {\bibfnamefont {M.}~\bibnamefont {Weides}}, \ and\ \bibinfo {author}
  {\bibfnamefont {A.~V.}\ \bibnamefont {Ustinov}},\ }\href@noop {} {\bibfield
  {journal} {\bibinfo  {journal} {Nature communications}\ }\textbf {\bibinfo
  {volume} {8}},\ \bibinfo {pages} {779} (\bibinfo {year} {2017})}\BibitemShut
  {NoStop}%
\bibitem [{\citenamefont {Grimsmo}\ and\ \citenamefont
  {Parkins}(2013)}]{grimsmo2013cavity}%
  \BibitemOpen
  \bibfield  {author} {\bibinfo {author} {\bibfnamefont {A.~L.}\ \bibnamefont
  {Grimsmo}}\ and\ \bibinfo {author} {\bibfnamefont {S.}~\bibnamefont
  {Parkins}},\ }\href@noop {} {\bibfield  {journal} {\bibinfo  {journal}
  {Physical Review A}\ }\textbf {\bibinfo {volume} {87}},\ \bibinfo {pages}
  {033814} (\bibinfo {year} {2013})}\BibitemShut {NoStop}%
\bibitem [{\citenamefont {Li}\ \emph {et~al.}(2019)\citenamefont {Li},
  \citenamefont {Macridin},\ and\ \citenamefont {Spentzouris}}]{andy2019}%
  \BibitemOpen
  \bibfield  {author} {\bibinfo {author} {\bibfnamefont {A.~C.~Y.}\
  \bibnamefont {Li}}, \bibinfo {author} {\bibfnamefont {A.}~\bibnamefont
  {Macridin}}, \ and\ \bibinfo {author} {\bibfnamefont {P.}~\bibnamefont
  {Spentzouris}},\ }\href@noop {} {\bibfield  {journal} {\bibinfo  {journal}
  {manuscript in preparation}\ } (\bibinfo {year} {2019})}\BibitemShut
  {NoStop}%
\bibitem [{\citenamefont {Somma}\ \emph {et~al.}(2003)\citenamefont {Somma},
  \citenamefont {Ortiz}, \citenamefont {Knill},\ and\ \citenamefont
  {Gubernatis}}]{somma2003quantum}%
  \BibitemOpen
  \bibfield  {author} {\bibinfo {author} {\bibfnamefont {R.}~\bibnamefont
  {Somma}}, \bibinfo {author} {\bibfnamefont {G.}~\bibnamefont {Ortiz}},
  \bibinfo {author} {\bibfnamefont {E.}~\bibnamefont {Knill}}, \ and\ \bibinfo
  {author} {\bibfnamefont {J.}~\bibnamefont {Gubernatis}},\ }\href@noop {}
  {\bibfield  {journal} {\bibinfo  {journal} {International Journal of Quantum
  Information}\ }\textbf {\bibinfo {volume} {1}},\ \bibinfo {pages} {189}
  (\bibinfo {year} {2003})}\BibitemShut {NoStop}%
\bibitem [{\citenamefont {Geller}\ \emph {et~al.}(2015)\citenamefont {Geller},
  \citenamefont {Martinis}, \citenamefont {Sornborger}, \citenamefont
  {Stancil}, \citenamefont {Pritchett}, \citenamefont {You},\ and\
  \citenamefont {Galiautdinov}}]{geller2015universal}%
  \BibitemOpen
  \bibfield  {author} {\bibinfo {author} {\bibfnamefont {M.~R.}\ \bibnamefont
  {Geller}}, \bibinfo {author} {\bibfnamefont {J.~M.}\ \bibnamefont
  {Martinis}}, \bibinfo {author} {\bibfnamefont {A.~T.}\ \bibnamefont
  {Sornborger}}, \bibinfo {author} {\bibfnamefont {P.~C.}\ \bibnamefont
  {Stancil}}, \bibinfo {author} {\bibfnamefont {E.~J.}\ \bibnamefont
  {Pritchett}}, \bibinfo {author} {\bibfnamefont {H.}~\bibnamefont {You}}, \
  and\ \bibinfo {author} {\bibfnamefont {A.}~\bibnamefont {Galiautdinov}},\
  }\href@noop {} {\bibfield  {journal} {\bibinfo  {journal} {Physical Review
  A}\ }\textbf {\bibinfo {volume} {91}},\ \bibinfo {pages} {062309} (\bibinfo
  {year} {2015})}\BibitemShut {NoStop}%
\bibitem [{\citenamefont {Barkoutsos}\ \emph {et~al.}(2018)\citenamefont
  {Barkoutsos}, \citenamefont {Gonthier}, \citenamefont {Sokolov},
  \citenamefont {Moll}, \citenamefont {Salis}, \citenamefont {Fuhrer},
  \citenamefont {Ganzhorn}, \citenamefont {Egger}, \citenamefont {Troyer},
  \citenamefont {Mezzacapo} \emph {et~al.}}]{barkoutsos2018quantum}%
  \BibitemOpen
  \bibfield  {author} {\bibinfo {author} {\bibfnamefont {P.~K.}\ \bibnamefont
  {Barkoutsos}}, \bibinfo {author} {\bibfnamefont {J.~F.}\ \bibnamefont
  {Gonthier}}, \bibinfo {author} {\bibfnamefont {I.}~\bibnamefont {Sokolov}},
  \bibinfo {author} {\bibfnamefont {N.}~\bibnamefont {Moll}}, \bibinfo {author}
  {\bibfnamefont {G.}~\bibnamefont {Salis}}, \bibinfo {author} {\bibfnamefont
  {A.}~\bibnamefont {Fuhrer}}, \bibinfo {author} {\bibfnamefont
  {M.}~\bibnamefont {Ganzhorn}}, \bibinfo {author} {\bibfnamefont {D.~J.}\
  \bibnamefont {Egger}}, \bibinfo {author} {\bibfnamefont {M.}~\bibnamefont
  {Troyer}}, \bibinfo {author} {\bibfnamefont {A.}~\bibnamefont {Mezzacapo}},
  \emph {et~al.},\ }\href@noop {} {\bibfield  {journal} {\bibinfo  {journal}
  {Physical Review A}\ }\textbf {\bibinfo {volume} {98}},\ \bibinfo {pages}
  {022322} (\bibinfo {year} {2018})}\BibitemShut {NoStop}%
\bibitem [{\citenamefont {Avalle}\ and\ \citenamefont
  {Serafini}(2014)}]{avalle2014noisy}%
  \BibitemOpen
  \bibfield  {author} {\bibinfo {author} {\bibfnamefont {M.}~\bibnamefont
  {Avalle}}\ and\ \bibinfo {author} {\bibfnamefont {A.}~\bibnamefont
  {Serafini}},\ }\href@noop {} {\bibfield  {journal} {\bibinfo  {journal}
  {Physical Review Letters}\ }\textbf {\bibinfo {volume} {112}},\ \bibinfo
  {pages} {170403} (\bibinfo {year} {2014})}\BibitemShut {NoStop}%
\bibitem [{\citenamefont {Macridin}\ \emph
  {et~al.}(2018{\natexlab{a}})\citenamefont {Macridin}, \citenamefont
  {Spentzouris}, \citenamefont {Amundson},\ and\ \citenamefont
  {Harnik}}]{macridin2018digital}%
  \BibitemOpen
  \bibfield  {author} {\bibinfo {author} {\bibfnamefont {A.}~\bibnamefont
  {Macridin}}, \bibinfo {author} {\bibfnamefont {P.}~\bibnamefont
  {Spentzouris}}, \bibinfo {author} {\bibfnamefont {J.}~\bibnamefont
  {Amundson}}, \ and\ \bibinfo {author} {\bibfnamefont {R.}~\bibnamefont
  {Harnik}},\ }\href@noop {} {\bibfield  {journal} {\bibinfo  {journal}
  {Physical Review A}\ }\textbf {\bibinfo {volume} {98}},\ \bibinfo {pages}
  {042312} (\bibinfo {year} {2018}{\natexlab{a}})}\BibitemShut {NoStop}%
\bibitem [{\citenamefont {Macridin}\ \emph
  {et~al.}(2018{\natexlab{b}})\citenamefont {Macridin}, \citenamefont
  {Spentzouris}, \citenamefont {Amundson},\ and\ \citenamefont
  {Harnik}}]{macridin2018electron}%
  \BibitemOpen
  \bibfield  {author} {\bibinfo {author} {\bibfnamefont {A.}~\bibnamefont
  {Macridin}}, \bibinfo {author} {\bibfnamefont {P.}~\bibnamefont
  {Spentzouris}}, \bibinfo {author} {\bibfnamefont {J.}~\bibnamefont
  {Amundson}}, \ and\ \bibinfo {author} {\bibfnamefont {R.}~\bibnamefont
  {Harnik}},\ }\href@noop {} {\bibfield  {journal} {\bibinfo  {journal}
  {Physical Review Letters}\ }\textbf {\bibinfo {volume} {121}},\ \bibinfo
  {pages} {110504} (\bibinfo {year} {2018}{\natexlab{b}})}\BibitemShut
  {NoStop}%
\bibitem [{\citenamefont {Aleksandrowicz}\ \emph {et~al.}(2019)\citenamefont
  {Aleksandrowicz}, \citenamefont {Alexander}, \citenamefont {Barkoutsos},
  \citenamefont {Bello}, \citenamefont {Ben-Haim}, \citenamefont {Bucher},
  \citenamefont {Cabrera-Hern{\'a}dez}, \citenamefont {Carballo-Franquis},
  \citenamefont {Chen}, \citenamefont {Chen}, \citenamefont {Chow},
  \citenamefont {C{\'o}rcoles-Gonzales}, \citenamefont {Cross}, \citenamefont
  {Cross}, \citenamefont {Cruz-Benito}, \citenamefont {Culver}, \citenamefont
  {Gonz{\'a}lez}, \citenamefont {Torre}, \citenamefont {Ding}, \citenamefont
  {Dumitrescu}, \citenamefont {Duran}, \citenamefont {Eendebak}, \citenamefont
  {Everitt}, \citenamefont {Sertage}, \citenamefont {Frisch}, \citenamefont
  {Fuhrer}, \citenamefont {Gambetta}, \citenamefont {Gago}, \citenamefont
  {Gomez-Mosquera}, \citenamefont {Greenberg}, \citenamefont {Hamamura},
  \citenamefont {Havlicek}, \citenamefont {Hellmers}, \citenamefont {Herok},
  \citenamefont {Horii}, \citenamefont {Hu}, \citenamefont {Imamichi},
  \citenamefont {Itoko}, \citenamefont {Javadi-Abhari}, \citenamefont
  {Kanazawa}, \citenamefont {Karazeev}, \citenamefont {Krsulich}, \citenamefont
  {Liu}, \citenamefont {Luh}, \citenamefont {Maeng}, \citenamefont {Marques},
  \citenamefont {Mart{\'\i}n-Fern{\'a}ndez}, \citenamefont {McClure},
  \citenamefont {McKay}, \citenamefont {Meesala}, \citenamefont {Mezzacapo},
  \citenamefont {Moll}, \citenamefont {Rodr{\'\i}guez}, \citenamefont
  {Nannicini}, \citenamefont {Nation}, \citenamefont {Ollitrault},
  \citenamefont {O'Riordan}, \citenamefont {Paik}, \citenamefont {P{\'e}rez},
  \citenamefont {Phan}, \citenamefont {Pistoia}, \citenamefont {Prutyanov},
  \citenamefont {Reuter}, \citenamefont {Rice}, \citenamefont {Davila},
  \citenamefont {Rudy}, \citenamefont {Ryu}, \citenamefont {Sathaye},
  \citenamefont {Schnabel}, \citenamefont {Schoute}, \citenamefont {Setia},
  \citenamefont {Shi}, \citenamefont {Silva}, \citenamefont {Siraichi},
  \citenamefont {Sivarajah}, \citenamefont {Smolin}, \citenamefont {Soeken},
  \citenamefont {Takahashi}, \citenamefont {Tavernelli}, \citenamefont
  {Taylor}, \citenamefont {Taylour}, \citenamefont {Trabing}, \citenamefont
  {Treinish}, \citenamefont {Turner}, \citenamefont {Vogt-Lee}, \citenamefont
  {Vuillot}, \citenamefont {Wildstrom}, \citenamefont {Wilson}, \citenamefont
  {Winston}, \citenamefont {Wood}, \citenamefont {Wood}, \citenamefont
  {W{\"o}rner}, \citenamefont {Akhalwaya},\ and\ \citenamefont
  {Zoufal}}]{Qiskit}%
  \BibitemOpen
  \bibfield  {author} {\bibinfo {author} {\bibfnamefont {G.}~\bibnamefont
  {Aleksandrowicz}}, \bibinfo {author} {\bibfnamefont {T.}~\bibnamefont
  {Alexander}}, \bibinfo {author} {\bibfnamefont {P.}~\bibnamefont
  {Barkoutsos}}, \bibinfo {author} {\bibfnamefont {L.}~\bibnamefont {Bello}},
  \bibinfo {author} {\bibfnamefont {Y.}~\bibnamefont {Ben-Haim}}, \bibinfo
  {author} {\bibfnamefont {D.}~\bibnamefont {Bucher}}, \bibinfo {author}
  {\bibfnamefont {F.~J.}\ \bibnamefont {Cabrera-Hern{\'a}dez}}, \bibinfo
  {author} {\bibfnamefont {J.}~\bibnamefont {Carballo-Franquis}}, \bibinfo
  {author} {\bibfnamefont {A.}~\bibnamefont {Chen}}, \bibinfo {author}
  {\bibfnamefont {C.-F.}\ \bibnamefont {Chen}}, \bibinfo {author}
  {\bibfnamefont {J.~M.}\ \bibnamefont {Chow}}, \bibinfo {author}
  {\bibfnamefont {A.~D.}\ \bibnamefont {C{\'o}rcoles-Gonzales}}, \bibinfo
  {author} {\bibfnamefont {A.~J.}\ \bibnamefont {Cross}}, \bibinfo {author}
  {\bibfnamefont {A.}~\bibnamefont {Cross}}, \bibinfo {author} {\bibfnamefont
  {J.}~\bibnamefont {Cruz-Benito}}, \bibinfo {author} {\bibfnamefont
  {C.}~\bibnamefont {Culver}}, \bibinfo {author} {\bibfnamefont {S.~D. L.~P.}\
  \bibnamefont {Gonz{\'a}lez}}, \bibinfo {author} {\bibfnamefont {E.~D.~L.}\
  \bibnamefont {Torre}}, \bibinfo {author} {\bibfnamefont {D.}~\bibnamefont
  {Ding}}, \bibinfo {author} {\bibfnamefont {E.}~\bibnamefont {Dumitrescu}},
  \bibinfo {author} {\bibfnamefont {I.}~\bibnamefont {Duran}}, \bibinfo
  {author} {\bibfnamefont {P.}~\bibnamefont {Eendebak}}, \bibinfo {author}
  {\bibfnamefont {M.}~\bibnamefont {Everitt}}, \bibinfo {author} {\bibfnamefont
  {I.~F.}\ \bibnamefont {Sertage}}, \bibinfo {author} {\bibfnamefont
  {A.}~\bibnamefont {Frisch}}, \bibinfo {author} {\bibfnamefont
  {A.}~\bibnamefont {Fuhrer}}, \bibinfo {author} {\bibfnamefont
  {J.}~\bibnamefont {Gambetta}}, \bibinfo {author} {\bibfnamefont {B.~G.}\
  \bibnamefont {Gago}}, \bibinfo {author} {\bibfnamefont {J.}~\bibnamefont
  {Gomez-Mosquera}}, \bibinfo {author} {\bibfnamefont {D.}~\bibnamefont
  {Greenberg}}, \bibinfo {author} {\bibfnamefont {I.}~\bibnamefont {Hamamura}},
  \bibinfo {author} {\bibfnamefont {V.}~\bibnamefont {Havlicek}}, \bibinfo
  {author} {\bibfnamefont {J.}~\bibnamefont {Hellmers}}, \bibinfo {author}
  {\bibfnamefont {{\L}.}~\bibnamefont {Herok}}, \bibinfo {author}
  {\bibfnamefont {H.}~\bibnamefont {Horii}}, \bibinfo {author} {\bibfnamefont
  {S.}~\bibnamefont {Hu}}, \bibinfo {author} {\bibfnamefont {T.}~\bibnamefont
  {Imamichi}}, \bibinfo {author} {\bibfnamefont {T.}~\bibnamefont {Itoko}},
  \bibinfo {author} {\bibfnamefont {A.}~\bibnamefont {Javadi-Abhari}}, \bibinfo
  {author} {\bibfnamefont {N.}~\bibnamefont {Kanazawa}}, \bibinfo {author}
  {\bibfnamefont {A.}~\bibnamefont {Karazeev}}, \bibinfo {author}
  {\bibfnamefont {K.}~\bibnamefont {Krsulich}}, \bibinfo {author}
  {\bibfnamefont {P.}~\bibnamefont {Liu}}, \bibinfo {author} {\bibfnamefont
  {Y.}~\bibnamefont {Luh}}, \bibinfo {author} {\bibfnamefont {Y.}~\bibnamefont
  {Maeng}}, \bibinfo {author} {\bibfnamefont {M.}~\bibnamefont {Marques}},
  \bibinfo {author} {\bibfnamefont {F.~J.}\ \bibnamefont
  {Mart{\'\i}n-Fern{\'a}ndez}}, \bibinfo {author} {\bibfnamefont {D.~T.}\
  \bibnamefont {McClure}}, \bibinfo {author} {\bibfnamefont {D.}~\bibnamefont
  {McKay}}, \bibinfo {author} {\bibfnamefont {S.}~\bibnamefont {Meesala}},
  \bibinfo {author} {\bibfnamefont {A.}~\bibnamefont {Mezzacapo}}, \bibinfo
  {author} {\bibfnamefont {N.}~\bibnamefont {Moll}}, \bibinfo {author}
  {\bibfnamefont {D.~M.}\ \bibnamefont {Rodr{\'\i}guez}}, \bibinfo {author}
  {\bibfnamefont {G.}~\bibnamefont {Nannicini}}, \bibinfo {author}
  {\bibfnamefont {P.}~\bibnamefont {Nation}}, \bibinfo {author} {\bibfnamefont
  {P.}~\bibnamefont {Ollitrault}}, \bibinfo {author} {\bibfnamefont {L.~J.}\
  \bibnamefont {O'Riordan}}, \bibinfo {author} {\bibfnamefont {H.}~\bibnamefont
  {Paik}}, \bibinfo {author} {\bibfnamefont {J.}~\bibnamefont {P{\'e}rez}},
  \bibinfo {author} {\bibfnamefont {A.}~\bibnamefont {Phan}}, \bibinfo {author}
  {\bibfnamefont {M.}~\bibnamefont {Pistoia}}, \bibinfo {author} {\bibfnamefont
  {V.}~\bibnamefont {Prutyanov}}, \bibinfo {author} {\bibfnamefont
  {M.}~\bibnamefont {Reuter}}, \bibinfo {author} {\bibfnamefont
  {J.}~\bibnamefont {Rice}}, \bibinfo {author} {\bibfnamefont {A.~R.}\
  \bibnamefont {Davila}}, \bibinfo {author} {\bibfnamefont {R.~H.~P.}\
  \bibnamefont {Rudy}}, \bibinfo {author} {\bibfnamefont {M.}~\bibnamefont
  {Ryu}}, \bibinfo {author} {\bibfnamefont {N.}~\bibnamefont {Sathaye}},
  \bibinfo {author} {\bibfnamefont {C.}~\bibnamefont {Schnabel}}, \bibinfo
  {author} {\bibfnamefont {E.}~\bibnamefont {Schoute}}, \bibinfo {author}
  {\bibfnamefont {K.}~\bibnamefont {Setia}}, \bibinfo {author} {\bibfnamefont
  {Y.}~\bibnamefont {Shi}}, \bibinfo {author} {\bibfnamefont {A.}~\bibnamefont
  {Silva}}, \bibinfo {author} {\bibfnamefont {Y.}~\bibnamefont {Siraichi}},
  \bibinfo {author} {\bibfnamefont {S.}~\bibnamefont {Sivarajah}}, \bibinfo
  {author} {\bibfnamefont {J.~A.}\ \bibnamefont {Smolin}}, \bibinfo {author}
  {\bibfnamefont {M.}~\bibnamefont {Soeken}}, \bibinfo {author} {\bibfnamefont
  {H.}~\bibnamefont {Takahashi}}, \bibinfo {author} {\bibfnamefont
  {I.}~\bibnamefont {Tavernelli}}, \bibinfo {author} {\bibfnamefont
  {C.}~\bibnamefont {Taylor}}, \bibinfo {author} {\bibfnamefont
  {P.}~\bibnamefont {Taylour}}, \bibinfo {author} {\bibfnamefont
  {K.}~\bibnamefont {Trabing}}, \bibinfo {author} {\bibfnamefont
  {M.}~\bibnamefont {Treinish}}, \bibinfo {author} {\bibfnamefont
  {W.}~\bibnamefont {Turner}}, \bibinfo {author} {\bibfnamefont
  {D.}~\bibnamefont {Vogt-Lee}}, \bibinfo {author} {\bibfnamefont
  {C.}~\bibnamefont {Vuillot}}, \bibinfo {author} {\bibfnamefont {J.~A.}\
  \bibnamefont {Wildstrom}}, \bibinfo {author} {\bibfnamefont {J.}~\bibnamefont
  {Wilson}}, \bibinfo {author} {\bibfnamefont {E.}~\bibnamefont {Winston}},
  \bibinfo {author} {\bibfnamefont {C.}~\bibnamefont {Wood}}, \bibinfo {author}
  {\bibfnamefont {S.}~\bibnamefont {Wood}}, \bibinfo {author} {\bibfnamefont
  {S.}~\bibnamefont {W{\"o}rner}}, \bibinfo {author} {\bibfnamefont {I.~Y.}\
  \bibnamefont {Akhalwaya}}, \ and\ \bibinfo {author} {\bibfnamefont
  {C.}~\bibnamefont {Zoufal}},\ }\href {\doibase 10.5281/zenodo.2562110}
  {\enquote {\bibinfo {title} {Qiskit: An open-source framework for quantum
  computing},}\ } (\bibinfo {year} {2019})\BibitemShut {NoStop}%
\bibitem [{\citenamefont {Chin}\ \emph {et~al.}(2011)\citenamefont {Chin},
  \citenamefont {Prior}, \citenamefont {Huelga},\ and\ \citenamefont
  {Plenio}}]{chin2011generalized}%
  \BibitemOpen
  \bibfield  {author} {\bibinfo {author} {\bibfnamefont {A.~W.}\ \bibnamefont
  {Chin}}, \bibinfo {author} {\bibfnamefont {J.}~\bibnamefont {Prior}},
  \bibinfo {author} {\bibfnamefont {S.~F.}\ \bibnamefont {Huelga}}, \ and\
  \bibinfo {author} {\bibfnamefont {M.~B.}\ \bibnamefont {Plenio}},\
  }\href@noop {} {\bibfield  {journal} {\bibinfo  {journal} {Physical Review
  Letters}\ }\textbf {\bibinfo {volume} {107}},\ \bibinfo {pages} {160601}
  (\bibinfo {year} {2011})}\BibitemShut {NoStop}%
\bibitem [{\citenamefont {Spall}\ \emph {et~al.}(1992)\citenamefont {Spall}
  \emph {et~al.}}]{spall1992multivariate}%
  \BibitemOpen
  \bibfield  {author} {\bibinfo {author} {\bibfnamefont {J.~C.}\ \bibnamefont
  {Spall}} \emph {et~al.},\ }\href@noop {} {\bibfield  {journal} {\bibinfo
  {journal} {IEEE transactions on automatic control}\ }\textbf {\bibinfo
  {volume} {37}},\ \bibinfo {pages} {332} (\bibinfo {year} {1992})}\BibitemShut
  {NoStop}%
\bibitem [{\citenamefont {Vlastakis}\ \emph {et~al.}(2015)\citenamefont
  {Vlastakis}, \citenamefont {Petrenko}, \citenamefont {Ofek}, \citenamefont
  {Sun}, \citenamefont {Leghtas}, \citenamefont {Sliwa}, \citenamefont {Liu},
  \citenamefont {Hatridge}, \citenamefont {Blumoff}, \citenamefont {Frunzio}
  \emph {et~al.}}]{vlastakis2015characterizing}%
  \BibitemOpen
  \bibfield  {author} {\bibinfo {author} {\bibfnamefont {B.}~\bibnamefont
  {Vlastakis}}, \bibinfo {author} {\bibfnamefont {A.}~\bibnamefont {Petrenko}},
  \bibinfo {author} {\bibfnamefont {N.}~\bibnamefont {Ofek}}, \bibinfo {author}
  {\bibfnamefont {L.}~\bibnamefont {Sun}}, \bibinfo {author} {\bibfnamefont
  {Z.}~\bibnamefont {Leghtas}}, \bibinfo {author} {\bibfnamefont
  {K.}~\bibnamefont {Sliwa}}, \bibinfo {author} {\bibfnamefont
  {Y.}~\bibnamefont {Liu}}, \bibinfo {author} {\bibfnamefont {M.}~\bibnamefont
  {Hatridge}}, \bibinfo {author} {\bibfnamefont {J.}~\bibnamefont {Blumoff}},
  \bibinfo {author} {\bibfnamefont {L.}~\bibnamefont {Frunzio}},  \emph
  {et~al.},\ }\href@noop {} {\bibfield  {journal} {\bibinfo  {journal} {Nature
  communications}\ }\textbf {\bibinfo {volume} {6}},\ \bibinfo {pages} {8970}
  (\bibinfo {year} {2015})}\BibitemShut {NoStop}%
\bibitem [{\citenamefont {Cross}\ \emph {et~al.}(2018)\citenamefont {Cross},
  \citenamefont {Bishop}, \citenamefont {Sheldon}, \citenamefont {Nation},\
  and\ \citenamefont {Gambetta}}]{cross2018validating}%
  \BibitemOpen
  \bibfield  {author} {\bibinfo {author} {\bibfnamefont {A.~W.}\ \bibnamefont
  {Cross}}, \bibinfo {author} {\bibfnamefont {L.~S.}\ \bibnamefont {Bishop}},
  \bibinfo {author} {\bibfnamefont {S.}~\bibnamefont {Sheldon}}, \bibinfo
  {author} {\bibfnamefont {P.~D.}\ \bibnamefont {Nation}}, \ and\ \bibinfo
  {author} {\bibfnamefont {J.~M.}\ \bibnamefont {Gambetta}},\ }\href@noop {}
  {\bibfield  {journal} {\bibinfo  {journal} {arXiv preprint arXiv:1811.12926}\
  } (\bibinfo {year} {2018})}\BibitemShut {NoStop}%
\end{thebibliography}%

% SM
% \newpage


\section*{Supplementary material}

# Supplemental Material for "Variational Quantum Simulation of Ultrastrong Light-Matter Coupling"

Agustin Di Paolo,[1] Panagiotis Kl. Barkoutsos,[2] Ivano Tavernelli,[2] and Alexandre Blais[1,3]

[1]*Institut quantique and Département de Physique,
Université de Sherbrooke, Sherbrooke J1K 2R1 QC, Canada*
[2]*IBM Research GmbH, Zurich Research Laboratory, Säumerstrasse 4, 8803 Rüschlikon, Switzerland.*
[3]*Canadian Institute for Advanced Research, Toronto, ON, Canada*
(Dated: September 18, 2019)

## I. POLARON TRANSFORMATION AND QUANTUM-CIRCUIT COMPILATION OF THE POLARON VARIATIONAL FORM

### A. Polaron transformation for the multimode quantum Rabi Hamiltonian

In this section, we study the effect of the polaron transformation [Eq. (4) of the main text] on the multimode Rabi Hamiltonian [$N = 1$ in Eq. (2) of the main text]. We first consider a slightly more general unitary of the form

$$P_{\boldsymbol{f}} = \exp\left[\sigma^x \sum_{k=1}^{M} f_k(a_k - a_k^\dagger)\right], \quad (1)$$

where $\{f_k\}$ are real parameters to be determined below. By transforming $H \to H'_{\boldsymbol{f}} = P_{\boldsymbol{f}}^\dagger H P_{\boldsymbol{f}}$, where $\boldsymbol{f} = (f_1, \ldots, f_M)$, we arrive at

$$H'_{\boldsymbol{f}}/\hbar = \frac{\omega'_q}{2}\sigma^z q^\dagger_{-\boldsymbol{f}} q_{\boldsymbol{f}} + \sum_{k=1}^{M} \omega_k a_k^\dagger a_k + g'_k \sigma^x (a_k + a_k^\dagger) \\ + E_{\boldsymbol{f}}. \quad (2)$$

Here, $\omega'_q = \omega_q e^{-2\boldsymbol{f}\cdot\boldsymbol{f}}$ is a renormalized frequency for the two-level atom, $g'_k = g_k - \omega_k f_k$ are new light-matter coupling parameters and $E_{\boldsymbol{f}} = \sum_{k=1}^{M} \omega_k f_k^2 - 2g_k f_k$ is an energy constant. We have moreover defined the operator $q_{\boldsymbol{f}} = \exp(2\sigma^x \sum_{k=1}^M f_k a_k)$.

The parameters $f_k$ in Eq. (2) can reduce the effective light-matter coupling strength $g'_k$ and thus make the groundstate of such a Hamiltonian closer to that of the noninteracting case. This, however, holds if the higher-order corrections to $\sigma^z$ from the operator $q_{-\boldsymbol{f}}^\dagger q_{\boldsymbol{f}}$, which mixes the $k$-modes, can be made small at the same time. This leads to a compromise for the best value of $\boldsymbol{f}$ that can be resolved by optimization of such a parameter [1, 2]. More precisely, assuming that the groundstate of Eq. (2) results close to vacuum, minimizing the groundstate energy $E_{\boldsymbol{f}} - \omega'_q/2$ leads to the optimal condition $f_k = g_k/(\omega_k + \omega'_q)$ and to the implicit equation

$$\omega'_q = \omega_q e^{-2\sum_{k=1}^{M}[g_k/(\omega_k + \omega'_q)]^2}, \quad (3)$$

for the renormalized frequency of the atom. We note that the difficulty of solving the scalar equation Eq. (3) does not scale with the number of bosonic modes under consideration. This fact deserves to be highlighted as the optimal value of $\boldsymbol{f}$ is used to initialize the optimizer before executing the variational quantum algorithm presented in the main text and in more details in Sect. I B. As discussed in Sect. I C, this is no longer true in the general case of $N > 1$, where the complexity of initialization scales exponentially with $N$.

We can gain an understanding of how the polaron transformation works by analyzing the behavior of approximate solutions to Eq. (3) at the boundaries of the range $1 \geq \omega'_q/\omega_q \geq 0$. Indeed, for small coupling strengths $g_k$, we expect $\omega'_q$ to differ only slightly from $\omega_q$ and thus to be able to approximate $\omega'_q/\omega_q \simeq 1 - \epsilon$, where $\epsilon \ll 1$. In contrast, for large coupling strengths compared to the system frequencies, we expect $\omega'_q$ to vanish exponentially and thus $\omega'_q/\omega_q \simeq \epsilon$ to hold. In the latter situation, the parameters of the polaron transformation approach the asymptotic scaling $f_k \simeq g_k/\omega_k$ leading to $g'_k \to 0$ (along with $\omega'_q \to 0$). This analysis indicates that the light-matter system effectively decouples in the polaron frame [see Eq. (2)], both in the strong and deep-strong coupling regimes. As demonstrated below, the disentangling capabilities of this transformation and the solution of Eq. (3) interpolate smoothly in the intermediate ultrastrong coupling regime, making this tool suitable for investigating the light-matter groundstate in a very broad range of parameters.

We confirm the intuition developed in the above paragraph by analyzing both the single- and two-mode quantum Rabi Hamiltonians in resonance conditions $\omega_q = \omega_k \equiv \omega$, for $g_k \equiv g \in [0, \omega]$. This is also the regime of parameters considered in the main text. Here, however, we are also interested in quantifying how well the state $P|\text{vac}\rangle$ approximates the groundstate of the multi-mode Rabi Hamiltonian in all coupling regimes. Fig. 1 (a) compares the expectation value $\langle H' \rangle_{\text{vac}}$ to the exact groundstate energy of the single-mode (left) and two-mode configurations (right). We observe that $\langle H' \rangle_{\text{vac}}$ follows closely the exact solution in the range $g/\omega \in [0, 2]$. This agreement is even improved for larger coupling strengths. Panels (b)-(d) show the effective atom frequency $\omega'_q$ and coupling strengths $g'_k$ along with the $f_k$ parameters of the polaron transformation. Solid lines correspond to the numerical optimization of $\omega'_q$, while dashed lines are obtained analytically from a series expansion of the right-hand side of Eq. (3) to second order

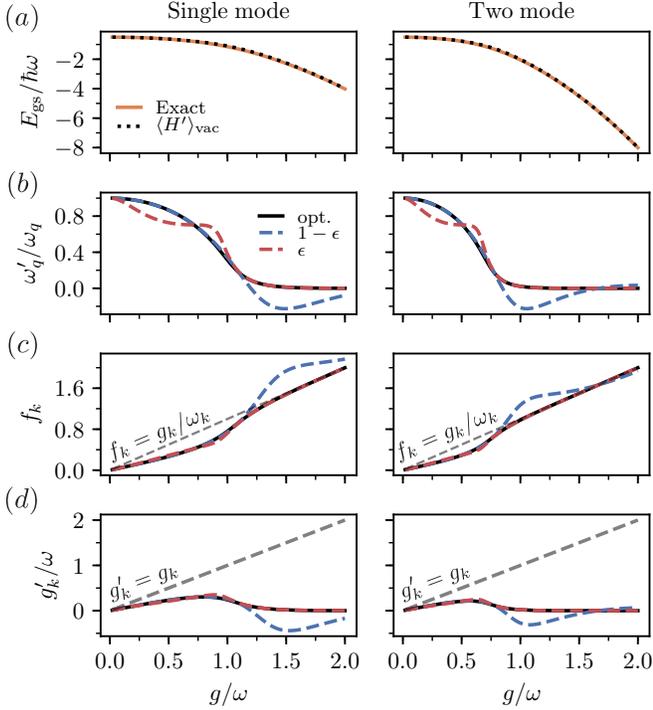

FIG. 1. Performance of polaron transformation in the strong-, ultrastrong- and deep-strong-coupling regimes for the single- and two-mode quantum Rabi models in resonance conditions. (a) Comparison between $\langle H' \rangle_{\text{vac}}$ and the exact groundstate energy. The present scale emphasizes the qualitative agreement between these two quantities, although it does not capture relatively small deviations which can be appreciated in the main text. Panels (b)-(d) show the polaron transformation parameters (black solid lines) obtained by numerical optimization. Dashed lines are analytical approximations of Eq. (3) at the boundaries of $1 \geq \omega_q'/\omega_q \geq 0$. The asymptotic regime of $f_k \simeq g_k/\omega_k$ is indicated in panel (c) (gray dashed lines). A similar line style in (d) indicates the coupling in the bare frame, $g_k$. We observe that $g_k'$ becomes exponentially small compared to $g_k$ as the bare coupling strength increases. This fact highlights the advantages of working in the polaron frame.

in the small parameter $\epsilon$ for both $\omega_q'/\omega_q \simeq 1 - \epsilon$ and $\omega_q'/\omega_q \simeq \epsilon$ regimes. As anticipated, we observe an exponential reduction of the ratio $\omega_q'/\omega_q$ along with the asymptotic tendency $f_k \simeq g_k/\omega_k$ and $g_k' \to 0$ as $g$ increases. Although the energy of the state $P|\text{vac}\rangle$ follows closely that of the groundstate in the laboratory frame in the DSC regime, we note that $f_k$ grows linearly with $g_k$ in such conditions. This unfavorable scaling would make the Trotter-decomposition-based approach employed in the main text rather inefficient, requiring quantum-circuits of larger depth to reach convergence.

The advantages of investigating the groundstate properties in the polaron frame are also emphasized by noticing that Eq. (2) can be recast into

$$H_f'/\hbar \simeq \frac{\omega_q'}{2}\sigma^z + \sum_{k=1}^{M} \omega_k a_k^\dagger a_k + 2\omega_q' f_k(\sigma^+ a_k + \sigma^- a_k^\dagger)$$
$$- 2\omega_q' \sigma^z \sum_{k,k'=1}^{M} f_k f_{k'} a_k^\dagger a_{k'} - \sum_{k=1}^{M} \frac{g_k^2}{\omega_k + \omega_q'} + \dots,$$
(4)

assuming $g_k'/(\omega_k + \omega_q') \ll 1$ and expanding $q_f$ to first order. Eq. (4) is reminiscent of the multimode Jaynes-Cummings model, with an additional term $\propto \sigma^z a_k^\dagger a_{k'}$, that can either shift the frequency of the atom as a function of the number of photons in the $k$-modes ($k = k'$) or allow for the exchange of an excitation between two $k$-modes through the atom ($k \neq k'$). Provided that the effect of the latter interaction is only perturbative, the groundstate of Eq. (4) is the vacuum state. As shown in the main text, these conditions also lead to relatively small $f_k$ parameters, making it possible to construct a short-depth variational form by Trotter decomposition of the polaron transformation.

## B. Quantum-circuit compilation of the polaron variational form

We now provide further details on the quantum circuit compilation of the polaron variational form designed for the multimode quantum Rabi model. Specifically, we seek to rewrite the unitary

$$\text{Varform} = \prod_{k=1}^{M} \prod_{s=1}^{d_k} \exp\left(\frac{f_k^s}{d_k}\sigma^x \widetilde{X}_k^e\right) \exp\left(\frac{f_k^s}{d_k}\sigma^x \widetilde{X}_k^o\right), \quad (5)$$

as a sequence of single- and two-qubit gates available on IBM's on-line quantum platform. The operator $\widetilde{X}_k^e$ in Eq. (5), which was introduced in the main text, is defined as

$$\widetilde{X}_k^e = -i \sum_{n_k \text{ even}}^{n_k^{\max}-1} \sqrt{n_k+1}(\sigma^x_{n_k}\sigma^y_{n_k+1} - \sigma^y_{n_k}\sigma^x_{n_k+1})/2. \quad (6)$$

$\widetilde{X}_k^o$ is defined analogously, with $n_k$ running over odds numbers in $[0, n_k^{\max} - 1]$. Since the Pauli products in Eq. (6) commute with each other, unitaries of the form $\exp(f_k^s \sigma^x \widetilde{X}_k^{e,o}/d_k)$ factorize into a sequence of controlled two-qubit gates that can be parallelized over the $k$-mode qubit registers. As shown in Fig. 2 (a), these two-qubit gates operate on pairs of next-neighbor qubits of the $k$-mode registers and are controlled by the atom qubit. Panel (b) provides the compilation of the controlled two-qubit gates in elementary single- and two-qubit gates [3].

The atom and the $k$-mode registers are initialized to the states $|0\rangle$ and $|\widetilde{0}_k\rangle = |10\dots 0\rangle$, respectively, which correspond to the noninteracting groundstates. This step

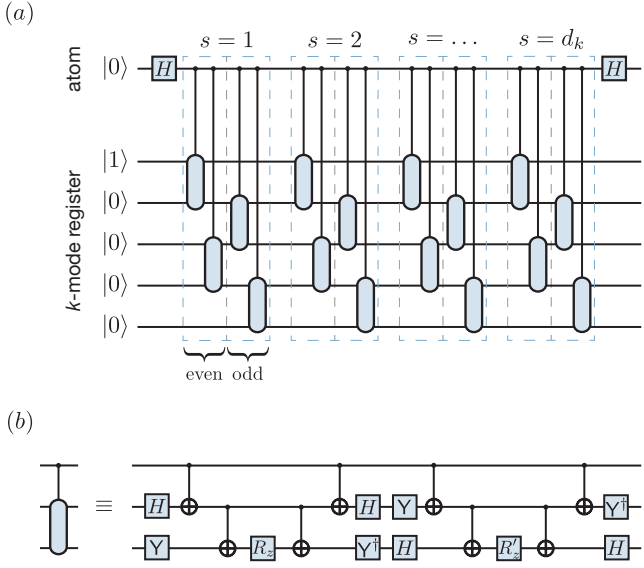

FIG. 2. Quantum-circuit compilation of the polaron variational form [Eq. (5)] for one of the $k$-mode registers. (a) Schematic representation of the variational form as a sequence of controlled two-qubit gates. Such gates correspond to the unitaries $\exp(f_k^s \sigma^x \widetilde{X}_k^e/d_k)$ and $\exp(f_k^s \sigma^x \widetilde{X}_k^o/d_k)$, performed at each Trotter step labeled by $s$. (b) Compilation of the controlled two-qubit gate in one- and two-qubit gates available on the IBM quantum hardware [3]. Here, $H$ denotes the Hadamard gate, $Y = R_x(\pi/2)$, and $R_z, R_z'$ are rotations around the $Z$ axis using the conventional notation in which $R_\mu(\theta) = \exp(-i\theta\sigma^\mu/2)$ for $\mu = x, y, z$. If the parameters $f_k^s$ are taken to be independent of the $k$-mode photon number, then $R_z = R_z(f_k^s \sqrt{n_k+1}/d_k)$ and $R_z' = R_z^\dagger$, where $n_k$ labels the sites within the $k$-mode register. If, in contrast, $f_k^s \to f_k^s(n_k)$ is allowed to vary from site to site, then $R_z = R_z[f_k^s(n_k)\sqrt{n_k+1}/d_k]$ and $R_z' = R_z^\dagger$—or $R_z' = R_z[f_k^{s'}(n_k)\sqrt{n_k+1}/d_k]$ if one wishes to introduce an extra degree of freedom, $f_k^{s'}(n_k)$, per controlled two-qubit gate. Importantly, the parameters of the variational form are initialized to the value specified by the polaron transformation in Eq. (4) of the main text.

is followed by $d_k$ sets of gates, where $d_k$ is the Trotter depth used to decompose the polaron unitary for mode $k$. In its simplest form, the polaron variational form has a single parameter per Trotter step. For a given $s$, the two-qubit-gate rotation angle is determined by such a parameter and the site index $n_k$ which introduces an additional scaling factor $\propto \sqrt{n_k}$ to this angle. The latter factor arises from the $\sqrt{n_k}$ scaling of the matrix elements of the harmonic-oscillator ladder operators ($a_k$ and $a_k^\dagger$) in the photon-number basis. A version of the variational form with a larger number of parameters can easily be crafted by letting the two-qubit-gate rotation angle vary and potentially depart from the $\sqrt{n_k}$ scaling. Moreover, an additional parameter can be introduced if the controlled two-qubit gates are implemented as in Fig. 2 (b), as this gate compilation uses two $R_z$ rotations that could be made independent from each other. In all cases, the initial value of the variational-form parameters must be set to that of the polaron transformation in Eq. (4) of the main text.

An alternative option to the gate in Fig. 2 (b) is the implementation a three-qubit gate at a hardware level without the need of gate compilation. This possibility, investigated in depth in Sect. V, has a number of advantages with respect to the compiled version of such gate. In fact, a hardware-level implementation of the controlled two-qubit gates would lead to a reduction of the gate count, potentially enabling the simulation of systems of larger size. Moreover, the gate generator can be engineered to conserve the number of excitations in the $k$-mode registers and thus the SES encoding. Since this criteria is not guaranteed by the complied version of the gate in Fig. 2 (b), an error occurring half-way in such a sequence could severely impact the state fidelity.

### C. Polaron transformation and variational form for the multimode Dicke Hamiltonian

In this section, we extend our approach to treat the multimode Dicke Hamiltonian [Eq. (2) of the main text]. To this end, the polaron transformation needs to be modified to include the $N$ atoms as

$$P_{\{\boldsymbol{f}_i\}} = \exp\left[\sum_{i=1}^{N}\sum_{k=1}^{M} \sigma_i^x f_{ik}(a_k - a_k^\dagger)\right], \quad (7)$$

where the real parameters $f_{ik}$, organized in the vectors $\boldsymbol{f}_i = (f_{i1},\ldots,f_{iM})$, for $i=1,\ldots,N$, now depend on both the atom and the $k$-mode indices [4]. By transforming the Hamiltonian in Eq. (2) of the main text as $H \to H'_{\{\boldsymbol{f}_i\}} = P_{\{\boldsymbol{f}_i\}}^\dagger H P_{\{\boldsymbol{f}_i\}}$, we find

$$\begin{aligned} H'_{\{\boldsymbol{f}_i\}}/\hbar = &\sum_{i=1}^{N} \frac{\omega'_{qi}}{2}\sigma_i^z q_{-\boldsymbol{f}_i}^\dagger q_{\boldsymbol{f}_i} + \sum_{k=1}^{M} \omega_k a_k^\dagger a_k \\ &+ \sum_{i=1}^{N}\sum_{k=1}^{M} g'_{ik}\sigma_i^x(a_k + a_k^\dagger) \\ &+ \sum_{i,i'=1}^{N} J_{ii'}\sigma_i^x \sigma_{i'}^x, \end{aligned} \quad (8)$$

which generalizes Eq. (2). Here, we derive parameters analogous to those introduced for the quantum Rabi model, including renormalized atom frequencies $\omega'_{qi} = \omega_{qi} e^{-2\boldsymbol{f}_i \cdot \boldsymbol{f}_i}$ and light-matter coupling constants $g'_{ik} = (g_{ik} - \omega_k f_{ik})$, along with the set of operators $q_{\boldsymbol{f}_i} = \exp(2\sigma_i^x \sum_{k=1}^{M} f_{ik} a_k)$ for $i=1,\ldots,N$. Unlike the $N=1$ case, Eq. (8) also includes an effective two-body coupling $J_{ii'} = \sum_{k=1}^{M}(\omega_k f_{ik} f_{i'k} - 2g_{ik} f_{i'k})$ between two atoms $i$ and $i'$, which is mediated by the $k$-modes.

Due to the latter interaction, the groundstate of Eq. (8) is not necessarily close to that of the noninteracting case. Instead, we assume a groundstate Ansatz of the form

$\prod_{k=1}^{M} |\psi_a\rangle|\widetilde{0}_k\rangle$, where $|\psi_a\rangle$ is the groundstate of the effective spin Hamiltonian

$$H_a/\hbar = \sum_{i=1}^{N} \frac{\omega'_{qi}}{2}\sigma_i^z + \sum_{i,i'=1}^{N} J_{ii'}\sigma_i^x\sigma_{i'}^x. \quad (9)$$

By minimizing the energy of $H_a$, one finds a parameterization $\{f_i\}$ of the polaron transformation, that approximately decouples the atoms from the $k$-modes in Eq. (8) [4]. In contrast to the case of the quantum Rabi model, the cost of finding a proper disentangling transformation now scales exponentially with system size.

Focusing on the limit of small $N$, Eq. (9) can still be handled on a classical processor which is used for diagonalizing $H_a$ and optimizing its groundstate energy as a function of $\{f_i\}$. It is worth mentioning that quantum routines, such as variational eigensolvers and quantum annealing, might also be useful for this task. In particular, the latter method is attractive for large $N$ due to the possibility of embedding Ising-type Hamiltonians with long-range interactions into physical models with bounded connectivity.

Regardless of how the above optimization is performed, a quantum circuit is needed to prepare the state $|\psi_a\rangle$ on the atom registers. This initialization step is followed by the application of the polaron variational form

$$\text{Varform} = \prod_{i=1}^{N}\prod_{k=1}^{M}\prod_{s=1}^{d_{ik}} \exp\left(\frac{f_{ik}^s}{d_{ik}}\sigma_i^x \widetilde{X}_k^e\right) \exp\left(\frac{f_{ik}^s}{d_{ik}}\sigma_i^x \widetilde{X}_k^o\right), \quad (10)$$

where $\sigma_i^x$ is the Pauli-$X$ operator for the $i^{\text{th}}$ atom, and $d_{ik}$ is the Trotter order of the polaron unitary involving this qubit and the $k$-mode labeled by $k$. Furthermore, $f_{ik}^s$ are the parameters of the polaron variational form, adding to a total of $\sum_{i=1}^{N}\sum_{k=1}^{M} d_{ik}$ parameters, which scales linearly with the number of atoms. Additional parameters can be introduced by allowing $f_{ik}^s \to f_{ik}^s(n_k)$ to depend on the $k$-mode photon-number index.

Fig. 3 shows a schematic of the variational form including the initialization step. We assume that $|\psi_a\rangle$ can be synthesized by a set of $d_a$ hardware-efficient layers acting on the atom registers [5]. Although this is not scalable to large $N$, the choice of a hardware-efficient approach is motivated by the lack of structure in Eq. (9). As previously shown in Fig. 2, the polaron variational form contains sets of controlled two-qubit gates acting on the $k$-mode registers. In the present case, the control qubit is swept across the atom registers, while the number of Trotter steps $d_{ik}$ may vary form one set to the other. Finally, we note that the compilation of the controlled two-qubit gates in one- and two-qubit gates is the same as in Fig. 2 (b).

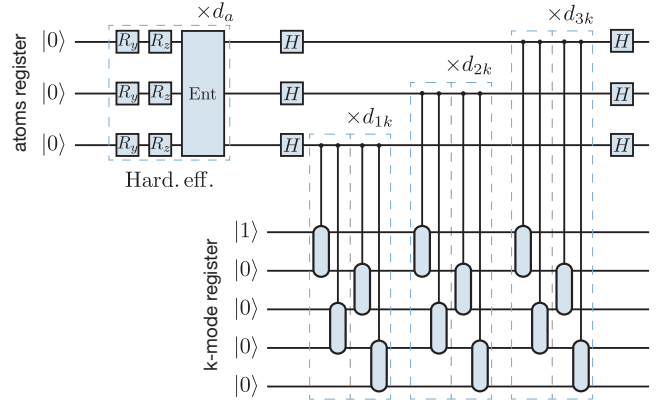

FIG. 3. Quantum circuit compilation of the variational form for the multimode Dicke problem for 3 atoms and a single $k$-mode register. The variational form is initialized on the state $\prod_{k=1}^{M} |\psi_a\rangle|\widetilde{0}_k\rangle$ by means of hardware-efficient layers acting on the atom registers. The hardware-efficient variational form is followed by layer of controlled two-qubit gates identical to those in Fig. 2 (b). Here, the control qubit is swept over the atom registers, and the number of Trotter steps may vary with the atom and $k$-mode indices.

## II. PERFORMANCE OF THE VQA ON A QUANTUM PROCESSOR

### A. VQA simulations with different hardware-noise levels

Mitigating the effect of noise is one of the greatest challenges for near term quantum computers. In order to first quantify this effect, it is useful to investigate the performance of the VQA for a variable noise strength. One way to modify the effective level of noise that a quantum algorithm is subject to, is to perform the quantum gates necessary for the computation having made these artificially slower. This enhances the effect of any decoherence channel and thus leads to an increased noise strength. Although we do not have low-level access to the pulses applied on the quantum hardware, like in Ref. [6], we can simulate the effect of a variable noise strength on the VQA by modifying the error model accordingly. This strategy also allow us to simulate a noise level below the calibrated values for the quantum device in use, which are provided by Qiskit [7].

We modify the noise level in simulation defining a noise factor, $\eta_{\text{noise}}$, such that

$$\begin{aligned} T_1 &= T_1^{\text{device}}/\eta_{\text{noise}} \\ T_2 &= T_2^{\text{device}}/\eta_{\text{noise}} \\ r_{\text{1q-g}} &= \eta_{\text{noise}}\, r_{\text{1q-g}}^{\text{device}} \\ r_{\text{2q-g}} &= \eta_{\text{noise}}\, r_{\text{2q-g}}^{\text{device}} \\ r_{\text{readout}} &= \eta_{\text{noise}}\, r_{\text{readout}}^{\text{device}}, \end{aligned} \quad (11)$$

where $T_1$ and $T_2$ are single-qubit relaxation and dephasing times, $r_{\text{1q-g}}$ and $r_{\text{2q-g}}$ are single- and two-qubit gate

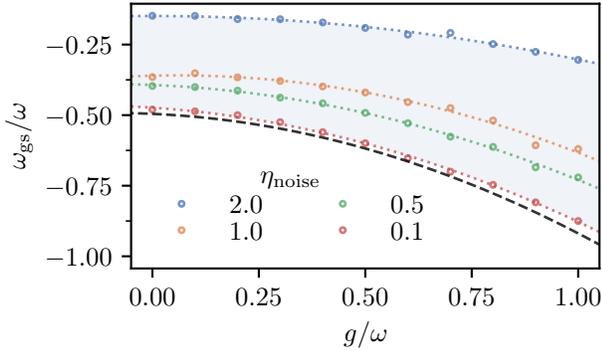

FIG. 4. Simulation of the VQA performance under variable noise strength. $\eta_{\text{noise}} = 1$ corresponds to the realistic noise model provided by Qiskit for the IBM Q Poughkeepsie device. The shared area indicates the range of results which are possible with levels of noise $\eta_{\text{noise}} = 0.1 - 2$. Extrapolation to zero-noise indicates that results with acceptable accuracy are only attainable with levels of noise one order of magnitude smaller than those in the current generation of devices.

error probabilities and $r_{\text{readout}}$ is the readout error probability, respectively. The quantities in Eq. (11) which are labeled as "device" correspond to calibrated values for the day that the runs were executed.

Fig. 4 shows the result of the simulations with variable noise strength. We use the Simultaneous-Perturbation-Stochastic-Approximation (SPSA) method as classical optimizer, with parameters $\alpha = 0.602$ and $\gamma = 0.101$ defined in Ref. [5]. We perform 25 calibration steps to compute the parameters $a$ and $c$ and another 100 SPSA trials for the actual optimization procedure. From the simulations, we conclude that VQA results with acceptable accuracy with respect to the exact groundstate energy would be attainable with noise levels that are one order of magnitude smaller than the actual ones. We note that extrapolation schemes to the zero-noise regime, like the ones discussed in Refs. [6, 8], can potentially help to further mitigate the effects of noise.

### B. Simulations on the quantum processor

For the quantum-hardware runs we use three qubits out of the twenty available on the IBM Q Poughkeepsie chip. The connectivity map of the device is shown in Fig. 5. The average error rates recorded throughout our experiments were $(5.25 \pm 0.212) \times 10^{-3}$ and $(3.75 \pm 0.364) \times 10^{-2}$ for single-quit gates and CNOT gates, respectively. These error rates are highly depended on the actual date on which the experiment took place. Mitigation of readout errors is done with the standardized methods provided by Qiskit-Ignis. There, a measurement calibration matrix is used to identify readout errors by preparing $2^N$ basis states and estimating the probability distribution of such state, with $N$ the number of qubits in simulation. The probability distribution of an unknown state can then be corrected based on these estimates. The calibration matrix was updated after every run or every 120 minutes of wall-clock time for the VQA runs.

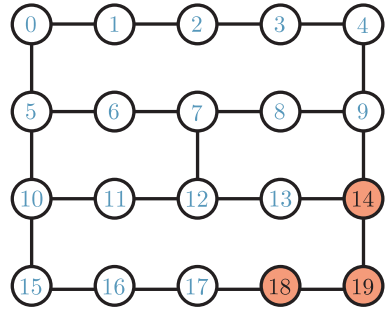

FIG. 5. Connectivity map of IBM Q Poughkeepsie quantum processor. Qubits 14, 18 and 19 (highlighted) were used for our experiment. The choice of qubits was depended on the respective single and two quit gate errors.

Beyond coherent and incoherent errors on quantum hardware, the main limitation to greater accuracy has been found to be the classical optimizer. Indeed, the SPSA method fails to acquire the expected solution in several cases, potentially getting stuck into local minima. We confirm this hypothesis indirectly by performing the optimization of the variational Ansatz in simulation assuming the calibrated noise model for the device. We then compute the expectation value of such variational states on quantum hardware. This additional experiment can reach significantly better accuracy than those obtained by running the optimization over quantum-hardware energy estimations, as shown in Fig. 2 of the main text. We therefore expect new and more powerful optimization algorithms to enable higher accuracy VQA results.

### III. SAMPLING THE JOINT WIGNER FUNCTION

We now provide details about the sampling of the joint Wigner distribution in Eq. (8) of the main text. We begin by noticing that measurements of the Pauli products $\sigma_1^{l_1} \ldots \sigma_N^{l_N}$ can be done in the computational basis, provided that a set of single-qubit gates $\{R_{l_i}\}$ are executed on the atom register prior to qubit readout. Taking this into consideration, we now focus on the case where the Pauli string $\sigma_1^{l_1} \ldots \sigma_N^{l_N}$ contains only $\sigma^z$ and/or identity operators and no prior rotation of $\rho$ is needed.

Eq. (8) of the main text makes use of the probability distribution of the displaced density matrix $\widetilde{D}^\dagger(\boldsymbol{\alpha})\tilde{\rho}\widetilde{D}(\boldsymbol{\alpha})$. In order to sample such a distribution, $\widetilde{D}^\dagger(\boldsymbol{\alpha})$ needs first to be compiled into single- and two-qubit gates. Since this joint-displacement operator is a product of single-mode displacements of the form

$\widetilde{D}^\dagger(\alpha_k)$, we only provide the quantum-circuit compilation for the latter unitary. To this end, it is convenient to introduce the real ($\alpha_k^R$) and imaginary ($\alpha_k^I$) parts of the displacement parameter $\alpha_k = \alpha_k^R + i\alpha_k^I$, and to expand the displacement operator as $\widetilde{D}^\dagger(\alpha_k) \simeq \exp[\alpha_k^R(\tilde{a}_k - \tilde{a}_k^\dagger)]\exp[-i\alpha_k^I(\tilde{a}_k + \tilde{a}_k^\dagger)]$, where "$\simeq$" indicates an equivalence up to a global phase. Making use of the site operators of the $k$-mode registers, we find

$$\tilde{a}_k - \tilde{a}_k^\dagger = -i\sum_{n_k=0}^{n_k^{\max}-1} \sqrt{n_k+1}(\sigma_{n_k}^x \sigma_{n_k+1}^y - \sigma_{n_k}^y \sigma_{n_k+1}^x)/2$$

$$\tilde{a}_k + \tilde{a}_k^\dagger = \sum_{n_k=0}^{n_k^{\max}-1} \sqrt{n_k+1}(\sigma_{n_k}^x \sigma_{n_k+1}^x - \sigma_{n_k}^y \sigma_{n_k+1}^y)/2. \quad (12)$$

Splitting the $k$-mode registers into even- and odd-index qubit subsets, the exponentiation of the operators in Eq. (12) can be implemented by a Trotter expansion, as it was done for the polaron variational form.

Fig. 6 summarizes the procedure for sampling the joint Wigner function with a set of tomography gates applied on the ultrastrong-coupling groundstate synthesized by the polaron variational form in panel (a). A first set of two-qubit gates, compiled in panel (b), implements a displacement operator along the imaginary-$\alpha_k$ axis with a Trotter order $d^I$. This is followed by a similar set of gates, compiled in panel (c), implementing a displacement along the real-$\alpha_k$ axis with Trotter depth $d^R$. Additionally, single-qubit tomography gates are applied on the atom registers. The circuit is terminated by readout of both the atom and $k$-mode registers. An histogram of counts (d) is constructed by repeating this procedure for a fixed ($\bm{l}, \bm{\alpha}$) pair. The joint Wigner function can then be computed from this histogram approximating the trace operator in Eq. (8) of the main text by

$$\text{Tr}_q[\ldots] \simeq \frac{2^M}{\pi^M}\sum_{\bm{q}}(-1)^{\sum_{i=1}^N \beta_{q_i}} c(q_1,\ldots,q_N;\tilde{n}_1,\ldots,\tilde{n}_M), \quad (13)$$

where $c(q_1,\ldots,q_N;\tilde{n}_1,\ldots,\tilde{n}_M)$ are the normalized counts for the basis vector $|q_1,\ldots,q_N;\tilde{n}_1,\ldots,\tilde{n}_M\rangle$ in which $q_i \in [0,1]$ is the state of the $i^{\text{th}}$ atom, and $\beta_{q_i} \in [0,1]$ accounts for the presence of a $\sigma_i^z$ operator before qubit readout. More precisely, such a parameter is set according to the rules $\beta_{q_i} = 1 - q_i$ if $l_i = z$, and $\beta_{q_i} = 0$ if $l_i = 0$. We note that the approximate relation in Eq. (13) can be replaced by an exact equivalence in the limit of large counts.

The reconstruction error scales with the amplitude of the displacement parameters $|\alpha_k|^2$, although it can be reduced by increasing the Trotter order in the implementation of the unitaries $\widetilde{D}^\dagger(\alpha_k)$. The finite Fock-state truncation of the encoded bosonic modes sets an upper bound to the accuracy of the reconstructed joint Wigner distribution, as demonstrated in Fig. 3 of the main text. Given that the quantum circuit corresponding to such

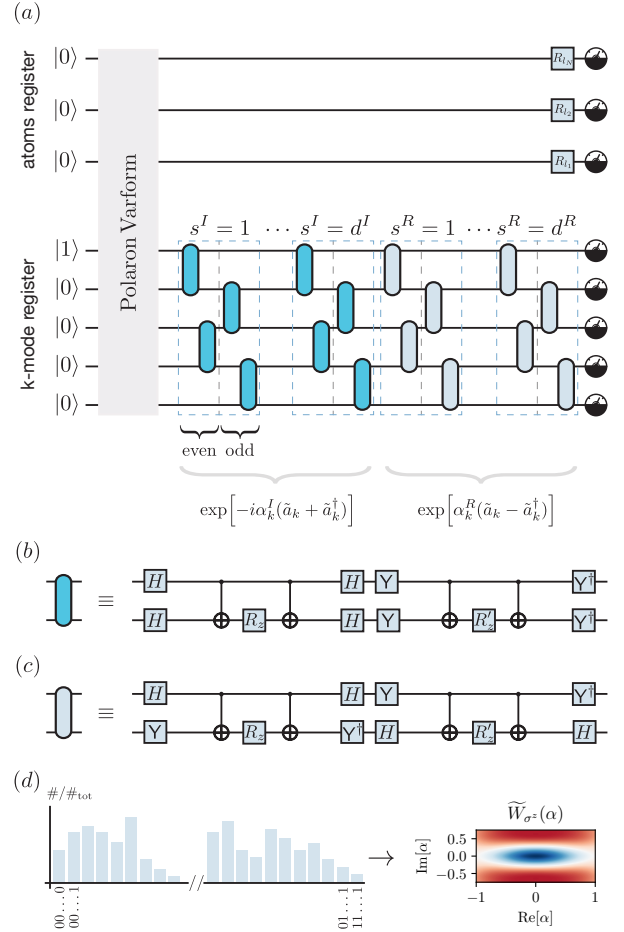

FIG. 6. Sampling the joint Wigner distribution corresponding to the atom and $k$-mode registers. (a) Quantum-circuit compilation of the displacement operators applied on the $k$-mode registers (here shown for a single mode), along with the single-qubit gates needed for tomography of the atom registers. These gates are executed on the ultrastrong-coupling groundstate synthesized by the polaron variational form. The $k$-mode displacement unitary is split in real and imaginary components, implemented by gate sequences with Trotter depth $d^R$ and $d^I$, respectively, which may in general be different. The imaginary component is depicted first, and makes use of the two-qubit gates in (b) where $R_z = R_z(\alpha_k^I\sqrt{\tilde{n}_k+1}/d^I)$ and $R_z' = R_z^\dagger$. What follows is the implementation of the real component of the displacement operator, which executes the two-qubit gates in (c) where $R_z = R_z(\alpha_k^R\sqrt{\tilde{n}_k+1}/d^R)$ and $R_z' = R_z^\dagger$. Qubit readout is performed at the end of the quantum circuit in (a), leading to the histogram of counts in (d) after several repetitions of the experiment. This allows for reconstruction of the joint Wigner function as described in the main text.

operators is appended to that of the polaron variational form, this procedure would ultimately be limited by the strength of the noise in the quantum processor. However, we find that Trotter depths as small as 2 are enough to demonstrate the qualitative features of the joint Wigner distribution.

## IV. EFFECT OF PHASE- AND BIT-FLIP NOISE CHANNELS UNDER A SES ENCODING

This section discuses the so-called memory error of a small qubit register encoding a bosonic mode. We consider both phase- and bit-flip error channels acting on a copy of the maximally entangled state $|\psi\rangle \sim \sum_{n_k=0}^{n_k^{\max}} |n_k\rangle$, in absence of logical gates. Specifically, we compute the state fidelity $F(\rho, \rho') = \text{Tr}[\sqrt{\sqrt{\rho}\rho'\sqrt{\rho}}]$ [9], where $\rho = |\psi\rangle\langle\psi| \otimes |\psi\rangle\langle\psi|$ and $\rho' \sim \sum_{n_k,n_k'=0}^{n_k^{\max}} |n_k\rangle\langle n_k'| \otimes \mathcal{E}_C(|n_k\rangle\langle n_k'|)$. Here, $\mathcal{E}_C$ are multiqubit error channels obtained by composition of single-qubit ones, $\mathcal{E}_{q,C}$. The latter have the general form $\mathcal{E}_{q,C}(\bullet) = \sum_{i=0}^{1} E_i \bullet E_i^\dagger$, where $\{E_i\}$ are the Kraus operator for the channel C. Denoting the error probability with $r$, we define the phase-flip channel by $E_0 = \sqrt{1-r}\mathbb{1}$, $E_1 = \sqrt{r}\sigma^z$, while $E_0 = \sqrt{1-r}\mathbb{1}$, $E_1 = \sqrt{r}\sigma^x$ correspond to the bit-flip channel.

Fig. 7 shows the result for the state fidelity assuming a $k$-mode register containing up to a maximum of 7 qubits. As anticipated in the main text, we find that bit-flip errors are the most relevant as the size of the $k$-mode register is increased. This can be understood intuitively by looking at the complement $\overline{\text{SES}}$ of the SES subspace used for the encoding. Since the number of basis vectors in $\overline{\text{SES}}$, and thus the dimension of this subspace, grows exponentially with $n_k^{\max}$, a noise operator breaking the SES symmetry could significantly affect the state fidelity in the limit of large $n_k^{\max}$. On the other hand, it is worth noticing that this might not necessarily limit the performance of near-term algorithms requiring only a small number of qubits. Alternatives for scaling-up to larger devices include the use of qubits with naturally long $T_1$ times, or a different encoding for the bosonic modes [10, 11]. Future work will investigate the performance crossover of the various possible encodings as the variational circuit is scaled up. Finally, we note that state fidelity, although standard, is a strong metric to evaluate the performance of our variational algorithm, and provides only a qualitative estimation of the impact on the energy of the variational ansatz.

## V. CIRCUIT-QED IMPLEMENTATION OF THE CONTROLLED-EXCHANGE GATES

With the purpose of reducing the gate count of the polaron variational form, we now present a superconducting-qubit implementation of a controlled-exchange gate. We stress, however, that the proposed approach could be leveraged by any other quantum-hardware platform with native interactions similar to those found in a standard circuit-QED setup. Below, we provide an ideal implementation of the gate interaction and then suggest a superconducting circuit that approaches the ideal scheme.

We first consider the case of a single atom and $k$-mode

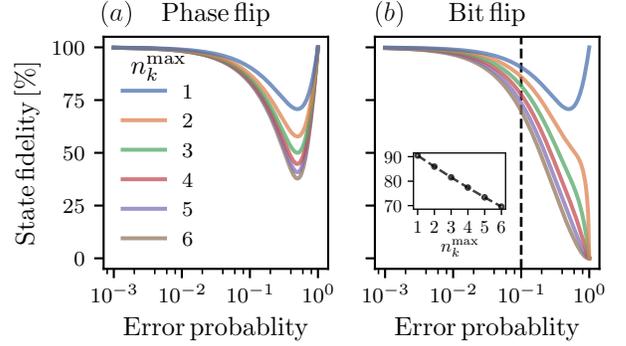

FIG. 7. Effect of phase-flip (a) and bit-flip (b) error channels on the fidelity of a maximally entangled state as a function of the total error probability. We consider a $k$-mode qubit register of size $n_k^{\max} + 1 \leq 7$. The legend applies for both left and right plots. We observe a significant decrease of the state fidelity as the size of the qubit register is increased. As discussed in the main text, bit-flip errors are expected to be dominant as the register size is scaled up to simulate a larger number of modes with also greater Fock-state truncation. The inset shows the state fidelity as a function of $n_k^{\max}$ for an error probability of $10^{-1}$.

registers. The frequency of the *physical* qubit corresponding to the atom is denoted by $\omega_\nu$, while the frequencies of the qubits belonging to the $k$-mode register are denoted by $\omega_{n_k}$, with $n_k \in [0, 1, \ldots, n_k^{\max}]$. Note that these frequencies are not related to the parameters of the problem that one wishes to simulate. With the purpose of engineering a controlled two-qubit gate, we assume the two qubits of the $k$-mode register–labeled by $\mu \in [n_k, n_k+1]$–to be independently coupled to the atom qubit with a time-dependent interaction strength. This situation is described by a 3-qubit Hamiltonian of the form

$$H_{\text{ideal}}/\hbar = \frac{\omega_\nu}{2}\sigma_\nu^z + \sum_\mu \left[\frac{\omega_\mu}{2}\sigma_\mu^z + \Omega_\mu(t)(\sigma_\mu^+\sigma_\nu^- + \sigma_\mu^-\sigma_\nu^+)\right] \tag{14}$$

where we take, in particular, $\Omega_\mu = \Omega_\mu^0 + 2\varepsilon_\mu \sin[(\omega_\mu - \omega_\nu + \delta_\mu)t + \phi_\mu]$. Here, $\Omega_\mu^0$ is an always-on interaction strength, $2\varepsilon_\mu$ is the modulation amplitude, $\delta_\mu$ is a frequency detuning with respect to the $\mu$-$\nu$ transition and $\phi_\mu$ is a relative phase. Counter-rotating terms of the form $\sigma_\mu^+\sigma_\nu^+$ and its Hermitian conjugate have been omitted after a RWA.

To make the three-qubit-gate interaction explicit, we now perform a standard time-dependent Schrieffer-Wolff transformation with generator

$$S = \sum_\mu \left[\frac{\Omega_\mu^0}{\omega_\mu - \omega_\nu} + \frac{i\varepsilon_\mu e^{-i[(\omega_\mu-\omega_\nu+\delta_\mu)t+\phi_\mu]}}{2(\omega_\mu-\omega_\nu)+\delta_\mu}\right]\sigma_\mu^+\sigma_\nu^- - \text{H.c.}, \tag{15}$$

conceived to remove first-order interaction terms by the condition $H_0 + [H_0, S] + H_{\text{int}} - i\dot{S} = 0$, where $H_0 = \frac{\omega_\nu}{2}\sigma_\nu^z + \sum_\mu \frac{\omega_\mu}{2}\sigma_\mu^z$ and $H_{\text{int}} = \sum_\mu \Omega_\mu(t)(\sigma_\mu^+\sigma_\nu^- + \sigma_\mu^-\sigma_\nu^+)$. Assuming $\Omega_\mu^0/(\omega_\mu - \omega_\nu) \ll 1$ (dispersive regime)

and $\varepsilon_\mu/[2(\omega_\mu - \omega_\nu) + \delta_\mu] \ll 1$, we expand the transformed Hamiltonian up to second order in the interaction strength, and move to a frame rotating at frequencies $\omega_\nu$ for the atom qubit and $\omega_\mu + \delta_\mu$ for the $k$-mode qubits, where the modulated interaction is resonant. Setting the phase of the drives as $\phi_{n_k} = 0$ and $\phi_{n_k+1} = -\pi/2$, and performing a second RWA, we find the effective Hamiltonian

$$H'_{\text{ideal}}/\hbar = \frac{\xi_{n_k}}{2}\sigma_\mu^z(\sigma_{n_k}^x \sigma_{n_k+1}^y - \sigma_{n_k}^y \sigma_{n_k+1}^x) + \frac{\delta\omega_\nu}{2}\sigma_\nu^z, \quad (16)$$

where drive parameters have been chosen to satisfy $-\delta_\mu + (\Omega_\mu^0)^2/(\omega_\mu - \omega_\nu) + \varepsilon_\mu^2/\delta'_\mu = 0$, with $\delta'_\mu = 2(\omega_\mu - \omega_\nu) + \delta_\mu$. The drive condition removes terms $\propto \sigma_\mu^z$ from Eq. (16) and makes the three-qubit interaction resonant in the current frame. Moreover, $\xi_{n_k}$ has been defined as an effective exchange-interaction rate between the two neighboring qubits of the $k$-mode register, that is mediated by the atom qubit and given by

$$\xi_{n_k} = \frac{1}{2}\frac{\varepsilon_{n_k}\varepsilon_{n_k+1}}{\delta'_{n_k}\delta'_{n_k+1}}(\delta'_{n_k} + \delta'_{n_k+1}). \quad (17)$$

Additionally, we derive a shift to the frequency of the atom qubit given by $\delta\omega_\nu = -(\Omega_\mu^0)^2/(\omega_\mu - \omega_\nu) - \varepsilon_\mu^2/\delta'_\mu$, due to the interaction with the two other qubits and the presence of the drive.

Evolution under the Hamiltonian in Eq. (16) generates the desired controlled-exchange gate operation $\exp[-i(\xi_{n_k}t/2)\sigma_\mu^z(\sigma_{n_k}^x \sigma_{n_k+1}^y - \sigma_{n_k}^y \sigma_{n_k+1}^x)]$ which is key to the polaron variational form. Due to the presence of the term $\propto \delta\omega_\nu$ in Eq. (16), an unintentional $R_z$ rotation on the atom qubit needs to be corrected for by applying an additional single-qubit gate. Modulation amplitudes $\varepsilon_\mu/2\pi$ of the order of 10 MHz and typical values of $\delta'_\mu/2\pi$ of the order of the GHz lead to controlled-exchange rates $\xi_{n_k}/2\pi$ in the range $0.1 - 0.5$ MHz. Despite this number being small compared to standard rates of one- and two-qubit gates in superconducting-qubit architectures, counting with a direct implementation of the three-qubit gate still provides a significant advantage with respect to its compiled counterpart in Fig. 2. In fact, the proposed gate is designed to conserve the excitation number of the $k$-mode registers and thus the SES encoding. Furthermore, while the gate time of the direct implementation is proportional to the desired rotation angle, the compiled version of the gate has an approximately fixed gate time determined by the number of CNOT gates in the circuit. This important difference would be leveraged further as the Trotter order of the polaron variational form is increased, making the controlled rotations closer to the identity.

Having presented an ideal model for the controlled-exchange gate, we now elaborate on a possible superconducting-circuit implementation of the Hamiltonian in Eq. (14). In particular, we consider an architecture made of transmon qubits and tunable couplers (see Fig. 8), similar to that studied in Ref. [12]. Using couplers to mediate parametric interactions allows us

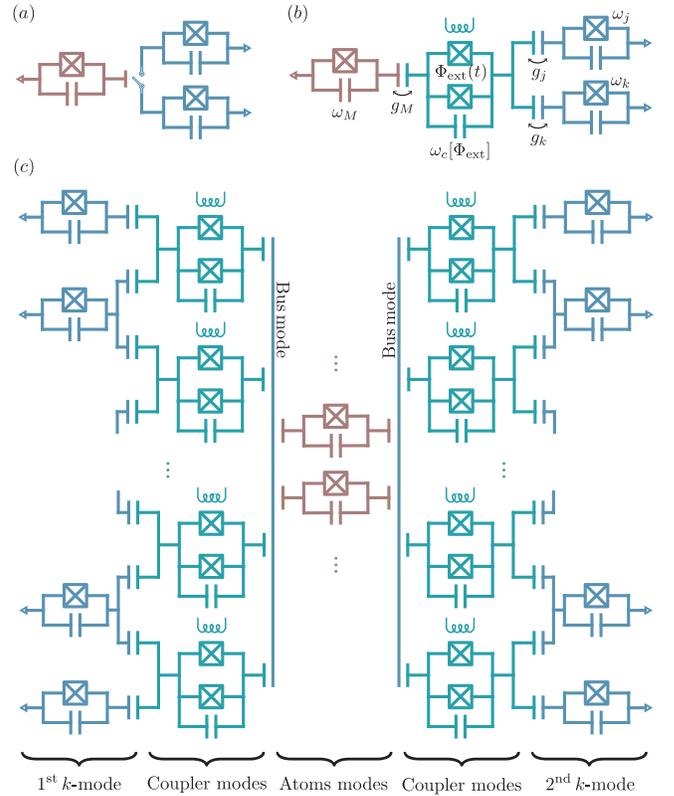

FIG. 8. Variational quantum-optics (VQO) superconducting processor. (a) Schematic of a controlled-exchange gate between three transmon qubits. The qubit in red plays the role of the atom register, controlling the switch on and off of an exchange interaction between two neighboring transmons belonging to the $k$-mode register. (b) Superconducting-qubit implementation of the concept in (a). A tunable-coupler (light green) is introduced to mediate the interaction between the atom and $k$-mode registers. (c) Device for the VQO simulation of the ultrastrong interaction between $N$ atoms (in red) and two bosonic modes (light blue, left and right). A superconducting resonator acting as a quantum bus is require to enable long-range interactions between the atoms and the coupler modes. Moreover, the bus mode enables dispersive two-qubit gates between the atom qubits if made frequency-tunable [13], which are required for state preparation in the case of $N > 1$.

to remove the need for frequency tunability of the qubit modes resulting in greater coherence times. Standard circuit quantization of the unit-cell device in Fig. 8 (b), followed by a two-level and rotating-wave approximations leads to the Hamiltonian

$$H = \frac{\omega_c^b[\Phi_{\text{ext}}]}{2}\sigma_c^z + \frac{\omega_\nu^b}{2}\sigma_\nu^z + \sum_\mu \frac{\omega_\mu^b}{2}\sigma_\mu^z \\ + g_\nu(\sigma_\nu^+\sigma_c^- + \sigma_\nu^-\sigma_c^+) + \sum_\mu g_\mu(\sigma_\mu^+\sigma_c^- + \sigma_\mu^-\sigma_c^+), \quad (18)$$

where $\omega_c^b[\Phi_{\text{ext}}]$ denotes the bare frequency of the tunable coupler, $\omega_\nu^b$ is the bare frequency of the atom qubit, $\{\omega_\mu^b\}$ are the bare frequencies of two neighboring qubits

in the $k$-mode register and $\{g_\nu, g_\mu\}$ are the respective coupling strengths between such qubits and the coupler. We assume that the coupler frequency can be tuned and modulated by a external magnetic flux $\Phi_{\text{ext}}$ through the coupler's SQUID loop.

Following [12], we perform the adiabatic elimination of the coupler mode by means of a Schrieffer-Wolff transformation in the dispersive regime $g_\nu \ll \Delta_\nu^{\text{b}}[\Phi_{\text{ext}}]$, $g_\mu \ll \Delta_\mu^{\text{b}}[\Phi_{\text{ext}}]$, where $\Delta_\beta^{\text{b}}[\Phi_{\text{ext}}] = \omega_\beta^{\text{b}} - \omega_c^{\text{b}}[\Phi_{\text{ext}}]$. Assuming that the coupler mode remains in its groundstate at all times, we derive an effective Hamiltonian of the form $H_{\text{eff}} = H_{\text{ideal}} + H_{\text{err}}$, where $H_{\text{ideal}}$ is the ideal interaction model given in Eq. (14) with frequency parameters $\omega_\nu = \omega_\nu^{\text{b}} + g_\nu^2/\Delta_\nu^{\text{b}}[\Phi_{\text{ext}}]$, $\omega_\mu = \omega_\mu^{\text{b}} + g_\mu^2/\Delta_\mu^{\text{b}}[\Phi_{\text{ext}}]$ and flux-tunable coupling strengths

$$\Omega_\mu[\Phi_{\text{ext}}(t)] = \frac{g_\mu g_\nu}{2} \frac{(\Delta_\mu^{\text{b}}[\Phi_{\text{ext}}] + \Delta_\nu^{\text{b}}[\Phi_{\text{ext}}])}{\Delta_\mu^{\text{b}}[\Phi_{\text{ext}}]\Delta_\nu^{\text{b}}[\Phi_{\text{ext}}]}. \quad (19)$$

$H_{\text{err}}$ is a spurious off-resonant term coupling directly the two qubits of the $k$-mode register. Implementation of the interaction model in Eq. (16) from $H_{\text{eff}}$ requires a two-tone modulation of $\Phi_{\text{ext}}(t)$ at frequencies $\omega_\mu - \omega_\nu + \delta_\mu$ for $\mu \in [n_k, n_k+1]$. We observe that the effect of $H_{\text{err}}$ can be exactly canceled by tuning the qubits to the destructive-interference condition $\Delta_{n_k}^{\text{b}}[\Phi_{\text{ext}}] = -\Delta_{n_k+1}^{\text{b}}[\Phi_{\text{ext}}]$. However, this also leads to a small interaction strength for the controlled-exchange operation. A better alternative is to consider the two $k$-mode qubits being coupled to an additional ancillary mode whose frequency is chosen to counteract the effect of $H_{\text{err}}$. Moreover, if the $k$-mode qubits are properly detuned the residual interaction only leads to a frequency renormalization of the drive condition above and to an off-resonant controlled-exchange interaction between $\nu$ and $n_k$ ($n_k + 1$) via $n_k + 1$ ($n_k$) which can be dropped by means of a RWA. As anticipated, we find that for typical circuit parameters and without optimization, the gate-interaction rate $\xi_{n_k}$ can reach values in the range of $0.1 - 0.5$ MHz, assuming a modulation amplitude between $25 - 50\%$ of $\Omega_\mu^0$ [12]. Further improvements on the speed of the gate might be enabled by optimization of the proposed circuit, the use of other possible coupling schemes implementing Eq. (16), or optimal control techniques [14].

The proposed implementation may be scaled-up to a larger number qubits, as shown schematically in Fig. 8 (c) for the case of $N$ atoms and two $k$-modes. A cavity bus mode is used to enable long-range interactions between the tunable couplers and the qubits playing the role of atoms registers. Moreover, the bus mode allows for the implementation of two-qubit gates between the atom qubits, which are necessary to initialize the polaron variational form for $N > 1$. We note that the controlled-exchange gates can be parallelized over even and odd qubits of the $k$-mode registers. Finally, we stress that scaling-up to a larger number of qubits entails issues that are beyond the scope of the present work and require to be examined in greater detail. The analysis of this section, however, provides a path forward towards the implementation of variational-quantum optics algorithms on special-purpose hardware.

---

[1] R. Silbey and R. A. Harris, The Journal of Chemical Physics **80**, 2615 (1984).
[2] T. Shi, Y. Chang, and J. J. García-Ripoll, Physical Review Letters **120**, 153602 (2018).
[3] J. D. Whitfield, J. Biamonte, and A. Aspuru-Guzik, Molecular Physics **109**, 735 (2011).
[4] G. Dìaz-Camacho, A. Bermudez, and J. J. García-Ripoll, Physical Review A **93**, 043843 (2016).
[5] A. Kandala, A. Mezzacapo, K. Temme, M. Takita, M. Brink, J. M. Chow, and J. M. Gambetta, Nature **549**, 242 (2017).
[6] A. Kandala, K. Temme, A. D. Córcoles, A. Mezzacapo, J. M. Chow, and J. M. Gambetta, Nature **567**, 491 (2019).
[7] G. Aleksandrowicz, T. Alexander, P. Barkoutsos, L. Bello, Y. Ben-Haim, D. Bucher, F. J. Cabrera-Hernádez, J. Carballo-Franquis, A. Chen, C.-F. Chen, J. M. Chow, A. D. Córcoles-Gonzales, A. J. Cross, A. Cross, J. Cruz-Benito, C. Culver, S. D. L. P. González, E. D. L. Torre, D. Ding, E. Dumitrescu, I. Duran, P. Eendebak, M. Everitt, I. F. Sertage, A. Frisch, A. Fuhrer, J. Gambetta, B. G. Gago, J. Gomez-Mosquera, D. Greenberg, I. Hamamura, V. Havlicek, J. Hellmers, L. Herok, H. Horii, S. Hu, T. Imamichi, T. Itoko, A. Javadi-Abhari, N. Kanazawa, A. Karazeev, K. Krsulich, P. Liu, Y. Luh, Y. Maeng, M. Marques, F. J. Martín-Fernández, D. T. McClure, D. McKay, S. Meesala, A. Mezzacapo, N. Moll, D. M. Rodríguez, G. Nannicini, P. Nation, P. Ollitrault, L. J. O'Riordan, H. Paik, J. Pérez, A. Phan, M. Pistoia, V. Prutyanov, M. Reuter, J. Rice, A. R. Davila, R. H. P. Rudy, M. Ryu, N. Sathaye, C. Schnabel, E. Schoute, K. Setia, Y. Shi, A. Silva, Y. Siraichi, S. Sivarajah, J. A. Smolin, M. Soeken, H. Takahashi, I. Tavernelli, C. Taylor, P. Taylour, K. Trabing, M. Treinish, W. Turner, D. Vogt-Lee, C. Vuillot, J. A. Wildstrom, J. Wilson, E. Winston, C. Wood, S. Wood, S. Wörner, I. Y. Akhalwaya, and C. Zoufal, "Qiskit: An open-source framework for quantum computing," (2019).
[8] K. Temme, S. Bravyi, and J. M. Gambetta, Physical Review Letters **119**, 180509 (2017).
[9] J. R. Johansson, P. D. Nation, and F. Nori, Computer Physics Communications **184**, 1234 (2013).
[10] A. Macridin, P. Spentzouris, J. Amundson, and R. Harnik, Physical Review Letters **121**, 110504 (2018).
[11] A. Macridin, P. Spentzouris, J. Amundson, and R. Harnik, Physical Review A **98**, 042312 (2018).
[12] D. C. McKay, S. Filipp, A. Mezzacapo, E. Magesan, J. M. Chow, and J. M. Gambetta, Physical Review Applied **6**, 064007 (2016).
[13] A. Blais, J. Gambetta, A. Wallraff, D. I. Schuster, S. M. Girvin, M. H. Devoret, and R. J. Schoelkopf, Physical Review A **75**, 032329 (2007).

[14] P. J. Liebermann, P.-L. Dallaire-Demers, and F. K. Wilhelm, arXiv preprint arXiv:1701.07870 (2017).



\end{document}